%% file: main.tex
\documentclass[letterpaper,twocolumn,10pt]{article}
\usepackage{usenix-2020-09}

\usepackage{tocloft}
\usepackage[frozencache,newfloat]{minted}
\usepackage{caption}

\newenvironment{code}{\captionsetup{type=listing}}{}
\SetupFloatingEnvironment{listing}{%
  name={Listing},
  fileext=lol}

\usepackage{listings}
\usepackage{amsmath}
\usepackage[ruled,vlined]{algorithm2e}
\usepackage{epsfig}
\usepackage{graphicx}
\usepackage{amssymb}
\usepackage{multirow}
\usepackage{pifont}
\usepackage{adjustbox}
\usepackage{array}
\usepackage{xpatch}
\usepackage[normalem]{ulem}
\usepackage{booktabs}
\usepackage{bigdelim}
\usepackage{colortbl}
\usepackage{url}

\usepackage{textcomp}
\newcommand{\textapprox}{\raisebox{0.5ex}{\texttildelow}}

\usepackage{lscape}

\usepackage{array}
\newcommand{\PreserveBackslash}[1]{\let\temp=\\#1\let\\=\temp}
\newcolumntype{C}[1]{>{\PreserveBackslash\centering}p{#1}}
\newcolumntype{R}[1]{>{\PreserveBackslash\raggedleft}p{#1}}
\newcolumntype{L}[1]{>{\PreserveBackslash\raggedright}p{#1}}
\newcolumntype{P}[1]{>{\centering\arraybackslash}p{#1}}

\usepackage{tikz}
\newcommand*\emptycirc[1][1ex]{\tikz\draw (0,0) circle (#1);}
\newcommand*\halfcirc[1][1ex]{%
  \begin{tikzpicture}
    \draw[fill] (0,0)-- (90:#1) arc (90:270:#1) -- cycle ;
    \draw (0,0) circle (#1);
\end{tikzpicture}}
\newcommand*\fullcirc[1][1ex]{\tikz\fill (0,0) circle (#1);}

\newtheorem{definition}{Definition}

\usepackage{xcolor}

\newcommand{\ie}{i.\@\,e.,\@\xspace}
\newcommand{\eg}{e.\@\,g.,\@\xspace}
\newcommand{\etal}{et~al.\@\xspace}

\newcommand{\pA}{\ding{182}\xspace}
\newcommand{\pB}{\ding{183}\xspace}
\newcommand{\pC}{\ding{184}\xspace}
\newcommand{\pD}{\ding{185}\xspace}

\newcommand{\algoName}{\textsc{StateInspector}\@\xspace}

\begin{document}

\date{}

\title{\Large \bf The Closer You Look, The More You Learn: \\ A Grey-box Approach to Protocol State Machine Learning}

\author{
{\rm Chris McMahon Stone}\\
University of Birmingham
\and
{\rm Sam L. Thomas}\\
University of Birmingham
\and
{\rm Mathy Vanhoef}\\
New York University Abu Dhabi
\and
{\rm James Henderson}\\
University of Birmingham
\and
{\rm Nicolas Bailluet}\\
ENS Rennes
\and
{\rm Tom Chothia}\\
University of Birmingham
}

\maketitle

\begin{abstract}
In this paper, we propose a new approach to infer state machine models from protocol implementations. Our method, \algoName, learns protocol states by using novel program analyses to combine observations of run-time memory and I/O. It requires no access to source code and only lightweight execution monitoring of the implementation under test. We demonstrate and evaluate \algoName's effectiveness on numerous TLS and WPA/2 implementations. In the process, we show \algoName enables deeper state discovery, increased learning efficiency, and more insightful post-mortem analyses than existing approaches. Further to improved learning, our method led us to discover several concerning deviations from the standards and a high impact vulnerability in a prominent Wi-Fi implementation.
\end{abstract}

\section{Introduction\label{sec:introduction}}

Flaws in protocol state machines have led to major vulnerabilities in many cryptographic protocols, ranging from TLS~\cite{de2015protocol} to WPA/2~\cite{vanhoef2017keyreinstallation}. Thus, the ability to extract these state machines from implementations and verify their correctness is an attractive means to perform security analysis. Moreover, since state machine models provide a succinct representation of the inner workings of an implementation---summarising tens of thousands of lines of code in a diagram that fits on a single page---they are also significantly easier to reason about and analyse than the implementation itself.
Current state-of-the-art approaches to extract state machines are built on active model learning algorithms that descend from the work of Angluin on learning regular sets~\cite{angluin1987learning}. Angluin introduced an algorithmic framework, the Minimally Adequate Teacher~(MAT), and the learning algorithm \textit{L}*. These techniques were later adopted for learning mealy-machine models~\cite{niese2003integrated,shahbaz2009inferring} and implemented in the open-source library LearnLib \cite{isberner2015open,raffelt2005learnlib}. The release of this library sparked a number of efforts to deploy model learning for applications such as conformance testing, legacy system inference, and most relevantly, security protocol analysis. In the security domain, it has been applied to various protocols, including TLS \cite{de2016tale,de2015protocol}, SSH \cite{SPIN17}, Wi-Fi \cite{stone2018extending}, TCP \cite{fiteruau2016combining}, OpenVPN \cite{daniel2018inferring}, and many others \cite{aarts2013formal,aarts2010inference,tappler2017model}.

An inherent limitation of these works, which all perform a type of \textit{black-box} model inference, is that they rely entirely on I/O observations. This means that any protocol behaviour which does not manifest via I/O, or does so beyond the preconfigured bound of the learning algorithm, cannot be extracted. In this paper, we present a new paradigm for model learning which overcomes this limitation. Our technique examines protocol behaviour ``under-the-hood'', with respect to an implementation's concrete execution state and how it handles I/O. As a result, we are able to learn models that capture more behaviour and obtain insights beyond what is possible by just considering I/O, allowing for easier model inspection and understanding.

Our state machine learning method works in two stages.
In the first stage, we capture snapshots of the implementation's execution context (\ie memory and registers) whilst it runs the protocol, which we then analyse to identify the locations in memory that define the protocol state for each possible state.
In the second stage, we learn the complete state machine by sending inputs to the protocol implementation and analysing the identified locations to see which values the state defining memory takes,
and so which protocol state has been reached.
This allows us to recognise each protocol state as soon as we reach it, making learning more effective than previous methods that require substantially more queries.

Similar to black-box learning, we are able to reason about the correctness of our approach, in the sense that we can list the assumptions that must hold for learning to terminate and result in a correct state machine.
We verify that our assumptions are reasonable and and realistic for analysing implementations of complex, widely deployed protocols, including TLS and WPA/2.
Further, we are able to demonstrate case studies of protocols that contain difficult to learn behaviour, that our method is able to learn correctly, but state-of-the-art black-box approaches cannot learn: either due to non-termination, or because they output the incorrect state machine.

\begin{table*}[t]
\footnotesize
\caption{Comparison of protocol model learning approaches. Style refers to active generation of inputs sequences, or passive trace replays.  Requirements denote: (P)rotocol Implementation, (A)bstraction functions for I/O, (T)races of protocol packets, and (F)uzzer. Learned information categorised as input (I), input to protocol state (I2PS), and input to concrete state (I2CS). \label{tbl:comparison}}
\centering
\begin{tabular}{
P{1cm}@{}
P{3cm}@{}
P{2.2cm}@{}
P{1.8cm}@{}
P{0.3cm}P{0.3cm}P{0.3cm}P{0.3cm}@{}
P{1.8cm}@{}
P{0.5cm}P{0.5cm}P{0.5cm}
P{2cm}}
\toprule
\textbf{} &
\textbf{} &
\multirow{2}{2.2cm}{\centering\textbf{Goals}} &
\multirow{2}{1.8cm}{\centering\textbf{Style}} &
\multicolumn{4}{c}{\textbf{Requirements}}
& \multirow{2}{1.8cm}{\centering\textbf{Crypto-protocols}}
& \multicolumn{3}{c}{\centering\textbf{What is learned?}}
& \multirow{2}{2cm}{\parbox{2cm}{\centering\textbf{State-classifier}}} \\
& & & & \textbf{P} & \textbf{A} & \textbf{T} & \textbf{F} & & \textbf{I} & \textbf{I2PS} & \textbf{I2CS} &  \\
\midrule
 \parbox[c]{2mm}{\multirow{3}{*}{\hspace{-0.25cm}\rotatebox[origin=c]{45}{\textbf{Black-box}}}} & \textsc{Model learning}~\cite{de2015protocol,stone2018extending} & \multirow{2}{*}{Model extraction} & \multirow{2}{*}{Active} & \multirow{2}{*}{\fullcirc} & \multirow{2}{*}{\fullcirc} & \multirow{2}{*}{\emptycirc} & \multirow{2}{*}{\emptycirc} & \multirow{2}{*}{\fullcirc} & \multirow{2}{*}{\emptycirc} & \multirow{2}{*}{\fullcirc} & \multirow{2}{*}{\emptycirc} & \multirow{2}{*}{I/O} \\
& \textsc{Pulsar}~\cite{gascon2015pulsar} & State-aware fuzzing & Passive & \emptycirc & \emptycirc & \fullcirc & \fullcirc & \emptycirc & \emptycirc & \fullcirc & \emptycirc & I/O \\
\midrule
\parbox[c]{2mm}{\multirow{5}{*}{\hspace{-0.25cm}\rotatebox[origin=c]{45}{\textbf{Grey-box}}}} &  \textsc{Prospex}~\cite{comparetti2009prospex} & Protocol RE & Passive & \emptycirc & \fullcirc & \fullcirc & \emptycirc & \emptycirc & \emptycirc & \fullcirc & \emptycirc & Control-Flow-I/O \\
& \textsc{MACE}~\cite{cho2011mace} & Input discovery & Active & \emptycirc & \fullcirc & \fullcirc & \emptycirc & \emptycirc & \fullcirc & \fullcirc & \emptycirc & I/O \\
& \textsc{AFLnet}~\cite{pham2020aflnet} & State-aware fuzzing & Active & \emptycirc & \fullcirc & \fullcirc & \fullcirc & \halfcirc & \fullcirc & \fullcirc & \emptycirc & I/O \\
& \textbf{\algoName} & \textbf{Model extraction} & \textbf{Active} & \fullcirc & \fullcirc & \emptycirc & \emptycirc & \fullcirc & \emptycirc & \fullcirc & \fullcirc & \textbf{Memory} \\
\bottomrule
\end{tabular}\\
\vspace{5pt}
\end{table*}

\bigskip

\noindent\textit{Contribution Summary:}
\begin{itemize}
    \item We present the design and implementation of a new model inference approach, \algoName, which leverages observations of run-time memory and program analyses. Our approach is the first which can map abstract protocol states directly to concrete program states.
    \item We demonstrate that our approach is able to learn states beyond the reach of black-box learning, while at the same time improve learning efficiency. Further, the models we produce allow for novel, post-mortem memory analyses not possible with other approaches.

    \item We evaluate our approach by learning the models of widely used TLS and WPA/2 implementations.
    Our findings include
    a high impact vulnerability, several concerning deviations from the standards, and unclear documentation leading to protocol misconfigurations.
\end{itemize}
To facilitate future research, we make \algoName and its test corpus publicly available as open source~\cite{github-url}.

We organise the paper as follows: we first provide the necessary background on model learning and binary analysis, then discuss the limitations of prior work and present a set of guiding research questions. Next, we present a high-level overview of our method, \algoName, and the assumptions we make, and follow it with a more in-depth discussion of our algorithm and learning framework. We then provide a thorough evaluation with respect to the current state of the art, and our research questions, for a number of TLS and WPA/2 implementations. Finally, we discuss the validity of our assumptions and the implications if they do not hold.

\section{Background}

\subsection{Model Learning}
\label{sec:modelformal}

The models we learn take the form of a \textit{Mealy machine}. For protocols, this type of model captures the states of the system which can be arrived at by sequences of inputs from a set $I$, and trigger corresponding outputs from a set $O$.
We consider Mealy machines that are \textit{complete} and \textit{deterministic}. This means that for each state, and input $i \in I$, there is exactly one output and one possible succeeding state.

Mealy machine learning algorithms include $L^*$~\cite{angluin1987learning,niese2003integrated,shahbaz2009inferring} and the more recent space-optimal TTT algorithm~\cite{isberner2014ttt}.
In essence, all of these algorithms work by systematically posing \textit{output queries} from $I^+$ (\ie non-empty strings of messages), each preceded by a special \textit{reset query}. These queries are mapped from abstract strings to concrete messages by an oracle, and then forwarded to the System Under Test (SUT). Each response from the SUT is then mapped back to an abstract output $o \in O$ for each prefix of the input message string, \ie $p \in I^+$. Once a hypothesis which achieves specific properties, namely \textit{completeness} and \textit{closedness}, has been built, a secondary procedure begins where \textit{equivalence} queries are posed to the teacher. These queries attempt to find counter-examples to refine the hypothesis. In black-box settings, these can be generated in various ways, including random testing, or more formally complete methods like Chow's W-Method \cite{chow1978testing}. Unfortunately, using the W-Method is highly expensive, essentially requiring an exhaustive search of all possible input sequences up to a specified bound. Naturally, a higher bound increases the number of queries required to learn a state machine and any states beyond this bound cannot be learned. In order to strike a trade-off between ensuring learning is tractable and obtaining representative model, selection of this bound requires careful consideration. Learned models can be analysed for interesting behaviour either manually, or automatically, \eg by using formal model checking~\cite{SPIN17,fiteruau2016combining}.

\subsection{Binary Program Analysis}

Our method does not require access to the source code of the SUT. Instead, we perform our analyses at the level of assembly language instructions, registers, and program memory. We base our analyses on two techniques: taint propagation and symbolic execution. Both methods compute properties of memory locations and registers along an execution path.

Taint propagation tracks the influence of \emph{tainted} variables by instrumenting code and tracking data at run time.
Any new value that is derived from a tainted value will also carry a taint, for example, if variable~\texttt{A} is tainted and influences variable~\texttt{B}, and variable~\texttt{B} in turn influences \texttt{C}, then variable \texttt{C} is also tainted meaning it is (indirectly) influenced by~\texttt{A}.

Symbolic execution computes constraints over \emph{symbolic} variables based on the operations performed on them.
For instance, suppose \texttt{A} is marked as symbolic, and a variable \texttt{B} is derived from \texttt{A} using some expression \verb|B=f(A)|, then symbolic execution will track this relation between both variables. More importantly, if we encounter a branch condition (\eg \texttt{if-then-else}) that depends on a symbolic variable, we can query an SMT solver to obtain suitable assignments that explore all feasible branches.

\section{Motivation\label{sec:motivation} and Related Work}

In Table \ref{tbl:comparison}, in addition to \algoName, we summarise five key approaches whose central aim is to learn a model of the state machine implemented by a protocol.
Each approach learns this model for different reasons (\ie fuzzing, reverse-engineering, or conformance testing), and their assumptions also differ (\ie black or grey-box access, known or unknown target protocols, and active or static trace-based learning).
In this section, we discuss the fundamental limitations of these key approaches, and demonstrate the need for a new direction. In doing so, we relate other works whose contributions also bear noteworthy relevance. We finish by presenting a set of research questions that serve as guiding principles for the design and evaluation of our proposal.

\smallskip
\noindent\textbf{Limited Coverage}
Learned model coverage in passive trace-based learning approaches (\eg \cite{gascon2015pulsar,comparetti2009prospex}) is entirely determined by the quality of pre-collected packet traces, and consequently is most appropriate for learning unknown protocols.
On the other hand, since active methods can generate protocol message sequences on-the-fly, their coverage capability is self-determined.
All state-of-the-art active methods (black-box~\cite{de2015protocol,stone2018extending,de2016tale,fiterau-brostean-usenix2020,SPIN17} and grey-box~\cite{cho2011mace,pham2020aflnet}) use automata learning algorithms (\eg L* or TTT) for their query generation and model construction, and such algorithms are known to have two key limitations. First, queries to a target system are constructed with a fixed, pre-defined set of inputs (\ie protocol messages). This means that behaviour triggered with inputs outside of this set cannot be explored. To help alleviate this issue, two approaches integrating automata learning with grey-box analyses have been proposed---MACE~\cite{cho2011mace}, which uses concolic execution, and AFLnet~\cite{pham2020aflnet}, which uses fuzzing. Despite some success when applied to text-based protocols, these and other existing model learning techniques suffer from a second issue, namely, limited exploration with respect to the \emph{depth} of the state machine.

\begin{figure}[t]
\centering
\resizebox{0.9\linewidth}{!}{
\footnotesize
\def\svgwidth{1.15\linewidth}
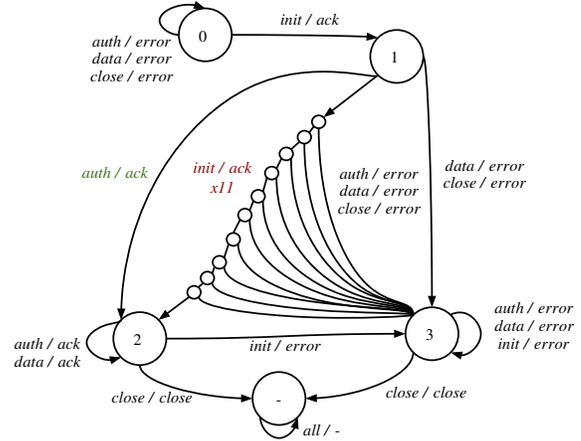
}
\caption{
State machine with a deep-state backdoor that can only be activated by sending 12 consecutive \textit{init} messages.
}
\label{fig:blackbox-fail}
\end{figure}

Consider the state machine in Figure~\ref{fig:blackbox-fail}. In this simple protocol, a \textit{backdoor} exists that allows us to transition from state 1 (not authenticated) to state 2 (authenticated) by supplying an input of 11 \texttt{init} messages, as opposed to an \texttt{auth} message\footnote{This backdoor is similar to a flaw found in a WPA/2 implementation \cite{stone2018extending}, which permitted a cipher downgrade after 3 failed steps of the handshake.}. To learn this state machine, state-of-the-art model learning with the TTT-Algorithm \cite{isberner2014ttt} and modified W-Method \cite{de2015protocol} requires \textapprox 360k queries to discover the backdoor, but fails to terminate after 1M queries---essentially operating by brute-force. This is because the number of queries posed is polynomial in the number of messages, states and exploration depth. Moreover, to identify this backdoor, we must set a maximum exploration depth of at least 11. In practice, however, since query complexity explodes with a high bound, most works use a much lower bound {(\eg 2 in ~\cite{de2015protocol,SPIN17})}, or use random testing which lacks completeness. This query explosion is also exacerbated by large input sets. Hence, for systems where time-spent-per-query is noticeable (\eg a few seconds, as is the case with some protocols), it is apparent that learning such systems requires a more efficient approach.

\smallskip
\noindent\textbf{Limited Optimisations} Similar to the problem of limited coverage, current approaches also lack sufficient information to optimise learning. That is, I/O observations do not enable learning effort to be focused where it is likely to be most fruitful. Consider de Ruiter et al.'s application of black-box learning to RSA-BSAFE for C TLS~\cite{de2015protocol}. Their learner cannot detect that the server does not close the connection upon handshake termination, and, thus, exhausts its exploration bound, posing tens of thousands of superfluous queries.

Prospex's grey-box approach~\cite{comparetti2009prospex} has the potential to avoid such an issue through its alternative modelling of states based on execution traces. However, it cannot capture states that only manifest as changes to memory, \eg counter increments per message read, and its coverage is inherently limited due to the use of trace-replays. Consequently, like MACE~\cite{cho2011mace} and AFLnet~\cite{pham2020aflnet}, it is not applicable to protocols with complex state or replay-protection (\ie security protocols).

\smallskip
\noindent\textbf{Limited Insight}
A further downside of methods that infer models based exclusively on I/O is that they may result in an over-abstraction. This can impede an analyst in obtaining a sufficiently detailed understanding of the implementation. Additionally, as active learning methods produce models that depend on how they map between concrete and abstract messages, their learned models will be inaccurate if the mapping is incorrect.
This mapping is done by a test harness, which is a highly flexible implementation of the target protocol (denoted by \textbf{P} in Table~\ref{tbl:comparison}).
The harness must be able to send and interpret protocol messages in an arbitrary order.
However, this is a complex task, and the developer of the harness often has to make (undocumented) decisions on how to handle unexpected or malformed messages.
As a result, accurately interpreting the resulting I/O model can be difficult to almost impossible, potentially making the resulting model effectively incorrect.
This downside is exemplified in a black-box analysis of TLS by de Ruiter \etal~\cite{de2015protocol}. In particular, the author's acknowledge that their state machines for OpenSSL 1.0.1g and 1.0.1j are incorrect due to differing assumptions between the implementation and their test harness.
Unfortunately, determining whether a resulting model is incorrect due to a flawed test harness, or whether there is in fact a bug in the tested implementation, is impossible based on the I/O model alone.
Instead, an expert must manually analyse both the harness and implementation to determine this.
Thus, I/O models alone provide limited insight, which extra grey-box information can help overcome.

We note that some works
extend black-box learning with taint analysis to learn register automata for protocols~\cite{howar2019combining,schrijvers2018learning}. Such automata are more insightful than Mealy machines, since they attempt to model conditions on state data for transitions between states. Unfortunately, these proposals assume access to source code and do not scale to \emph{real-world} protocols.

\smallskip
\noindent\textbf{Research Questions}
Based on the above limitations of current works,
we use the following research questions to guide the design of our new grey-box approach, \mbox{\algoName}:

\medskip
\noindent\textbf{RQ1:} Can we expand the capability of model learning to locate deep states typically beyond reach of current approaches, especially when learning with large input sets?

\medskip
\noindent\textbf{RQ2:} Can we exploit grey-box information in order to apply more targeted learning, resulting in efficiency gains?

\medskip
\noindent\textbf{RQ3:} Can we employ observations of internal SUT behaviour, \eg API usage or memory classifications of states, for more insightful analysis of the inferred I/O models?

\begin{figure*}[h!t]
  \centering
  \scalebox{0.9}{
  \footnotesize
  \def\svgwidth{0.9\linewidth}
  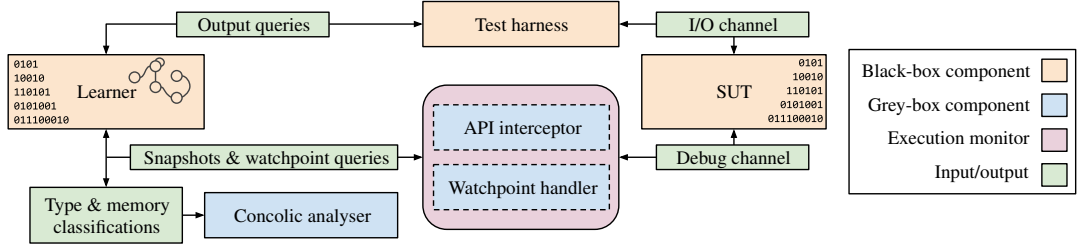}
  \caption{Grey-box State Learning Architecture.}
  \label{fig:arch}
\end{figure*}

\section{Overview}
\label{sec:overview}

We depict a high-level overview of our approach in Figure~\ref{fig:arch}. Our method combines insights into the runtime behaviour of the SUT (provided by the execution monitor and concolic analyser) with observations of its I/O behaviour (provided by the test harness) to learn models that provide comparable guarantees to traditional black-box approaches. We organise this section by first stating the assumptions we make about the state machines our approach learns, and then provide an overview of its operation. So that the reader has a clear understanding of all assumptions we make of the implementations we analyse, we also state assumptions about the configurable parameters of our method. We defer discussion of the choice of their concrete assignments, and how the assumptions can be loosened (and the implications of doing so) to later sections, so that they can be understood in context.

Our first assumption is standard for state machine learning:

\medskip
\noindent\textbf{Assumption 1.} \textit{The protocol state machine of the SUT is finite, and can be represented by a mealy machine.}
\medskip

In all concrete implementations, each state of this mealy machine will correspond to a particular assignment of values to specific ``state-defining'' memory locations, allocated by the implementation. We use two terms to discuss these locations and the values assigned to them, which we define as follows:

\begin{definition}[Candidate State Memory Location]
Any memory location that takes the same value after the same inputs, for any run of the protocol.
\end{definition}

\begin{definition}[A Set of State-Defining Memory]
\label{def:stateMemSet}
A minimal subset of candidate state memory locations whose values during a protocol run uniquely determine the current state.
\end{definition}

For simplicity, we also refer to a member of any state-defining memory set as \textit{state memory}. Our next assumption is that we know a priori the complete language needed to interact with the SUT and hence generate the state machine:

\medskip
\noindent\textbf{Assumption 2.} \textit{All protocol states can be reached via queries built up from the inputs known to our testing harness $I^+$.}
\medskip

Under these assumptions, we now describe the operation of our approach. Our learner uses a test harness to interact with the SUT. This test harness is specific to each protocol analysed and tracks session state, and is responsible for communicating with the SUT and translating abstract representations of input and output messages (for the learner) to concrete representations (for the SUT). In the first phase of learning, the learner instructs the test harness to interact with the SUT to exercise normal protocol runs, trigger errors, and induce timeouts multiple times. We capture snapshots of the SUT's execution context (\ie its memory) for each of these runs using an execution monitor. We perform subsequent analysis of the snapshots in order to determine a set of candidate memory locations that can be used to reason about which state the SUT is in. We make the following assumption about this memory:

\medskip
\noindent\textbf{Assumption 3.} \textit{A set of state-defining memory for all states is allocated along a normal protocol run, and this memory will take a non-default value in the normal run, during error handling, or timeout routines.}
\medskip

Any memory location found to have the same value for each sequence of inputs for each run is considered candidate state memory. And any location that takes the same value in all states is discarded. Assumption 3 ensures that this reduced set of locations will be a set of state-defining memory, however, this set may also contain superfluous locations. Further, this set of locations may not be state-defining for every state, and is rather a superset of locations for all states.

The next phase of learning uses the candidate set to construct a model of the protocol state machine. During this process, we identify the set of state-defining memory for each state. This allows us to determine if two states arrived at from the same input sequence should be considered equal, even if their memory differs, as we only need to consider the values of state-defining locations when performing this so-called equivalence check. It also allows us to effectively ignore any superfluous locations in our candidate set since they will not be considered state-defining for any state.

To learn the state machine, we queue all queries from $I^+$ to fulfill the \textit{completeness} property for states in a mealy machine. Namely, that all inputs are defined for each state. We perform these queries iteratively in increasing length. As in the first phase, we take snapshots of the execution state of the SUT on each I/O action. Each time a distinct state, according to its I/O behaviour and defined state memory, is discovered, we queue additional queries to ensure that the state is also fully defined (\ie we attempt to determine all states reachable from it). We continue building the automaton in this way until no new states are found.

As well as state memory, it is possible that our candidate set contains locations with assignments that appear to be state-defining for all states identified from the set of posed inputs, but are actually not. For instance, if a protocol implementation maintains a counter of the messages received, it might be mistaken for state memory, leading to a self loop in the state machine being conflated with an infinite progression of seemingly different states. Recognising such memory will prevent our framework from mistaking a looping state from an infinite series of states, and so ensures termination. To determine if such memory exists and if it can be ignored, we perform a \textit{merge check}. This allows us to replace a series of states repeated in a loop, with a single state and a self loop. This check is only performed between states that have the same observed input and output behaviour, and are connected at some depth (\ie one is reachable from the other). We make the following assumption about this depth:

\medskip
\noindent\textbf{Assumption 4.} \textit{The depth that we check for possible merge states is larger than the length of any loop in the state machine of the implementation being tested.}
\medskip

When performing a merge check, we consider two states equal if the values assigned to their state-defining locations are equal. Therefore, for each location that differs, we must determine if it behaves as state memory for each state (we discuss the conditions for this in Section~\ref{sec:uncertainmem}). To do so, we use a novel analysis (implemented in the concolic analyser in Figure~\ref{fig:arch}).
This analysis identifies a memory location as state-defining when this location influences decisions about which state we are in.
Concretely, this is encoded by checking if the location influences a branch that leads to a write to any candidate state memory. That is, state-defining memory will control the state by directly controlling writes to (other) state memory, and non-state-defining memory will not. This leads to our final assumption:

\medskip
\noindent\textbf{Assumption 5.} \textit{Any state-defining memory will control a branch, along an execution path triggered by messages from the set~~$I^+$, and it will lead to a write to candidate state memory within an a priori determined number of instructions.}
\medskip

We discuss the selection of the above bound in Section~\ref{sec:taint}, but note that it is a configurable parameter of our algorithm. We complete learning when no further states can be merged, and no new states (that differ in I/O behaviour or state memory assignment) can be discovered by further input queries. At this point, \algoName outputs a representative mealy machine for the SUT.

\section{Methodology}
\label{sec:algorithm}

The learner component of Figure~\ref{fig:arch} orchestrates our model inference process. It learns how the SUT interacts with the world by observing its I/O behaviour via the test harness, and learns how it performs state-defining actions at the level of executed instructions and memory reads and writes, via the execution monitor. Inference begins with a series of bootstrap queries in order to identify candidate state memory, \ie $M$. All queries posed to the SUT in this stage serve the dual purpose of identifying state memory and also refining the model. This ensures minimal redundancy and repetition of queries. Given an estimation of $M$, we then proceed to construct the state machine for the implementation by identifying each state by its I/O behaviour and set of state-defining memory. Throughout this process we perform \textit{merge checks} which determine if two states are equivalent with respect to the values of their state-defining locations.

\subsection{Monitoring Execution State\label{sec:monitoring-exec}}

To monitor a SUT's state while it performs protocol related I/O, we take snapshots of its memory and registers at carefully chosen points. As many protocols are time-sensitive, we ensure that this monitoring is lightweight to avoid impacting the protocol's behaviour. More specifically:
\begin{enumerate}
    \item We avoid inducing overheads such that the cost of each query becomes prohibitively high.
    \item We do not interfere with the protocol execution, by, for example, triggering timeouts.
    \item We only snapshot at points that enable us to capture each input's effect on the SUT's state.
\end{enumerate}

To perform a snapshot, we momentarily pause the SUT's execution and copy the value of its registers and contents of its mapped segments to a buffer controlled by the execution monitor. To minimise overheads, we buffer all snapshots in memory, and only flush them to disk after the SUT has finished processing a full query. We find that the overheads induced by snapshotting in this way are negligible and do not impact the behaviour of any of the protocols analysed.

To identify when to perform snapshots, we infer which system calls (syscalls) and fixed parameters a SUT uses to communicate with its client/server. We do this by generating a log of all system calls and passed parameters used during a standard protocol run. We then identify syscall patterns in the log and match them to provided inputs and observed outputs. The number of syscalls for performing I/O is small, hence the number of patterns we need to search for is also small, \eg combinations of \texttt{bind}, \texttt{recvmsg}, \texttt{recvfrom}, \texttt{send}, etc. When subsequently monitoring the SUT, we hook the identified syscalls and perform state snapshots when they are called with parameters matching the those in our log.

\subsubsection{Mapping I/O Sequences To Snapshots}
\label{sec:matchsnapshot}
\begin{figure}[h!]
  \centering
    \footnotesize
    \def\svgwidth{0.9\columnwidth}
    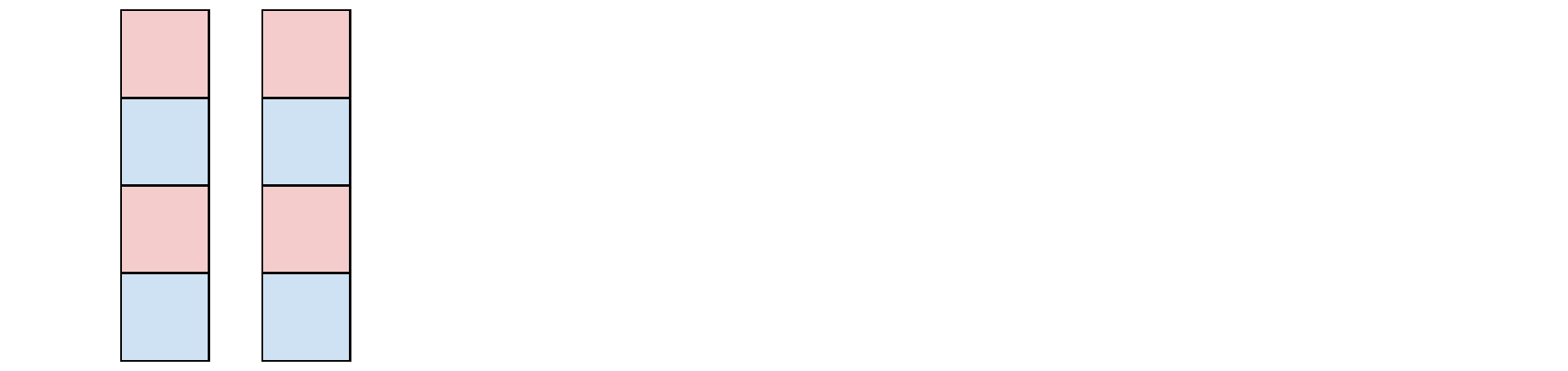
  \caption{Mapping I/O sequences to snapshots.
  We depict four cases \pA-\pD, snapshot events are shown on the left and I/O on the right.
  We indicate read/input events in red, write/output events in blue, state changes in yellow, and connection close events in green.
  \pA shows the case where snapshots and state changes are trivially aligned with I/O.
  \pB shows the case where a state change occurs without a corresponding output event.
  \pC shows the case where state changes occur after output events.
  \pD shows the case where a state change occurs after the last input event, with no output before the connection is closed.
  \label{fig:io-mapping}}
\end{figure}

To map I/O sequences to snapshots, we maintain a monotonic clock that is shared between our test harness (which logs I/O events) and our execution monitor (which logs snapshot events). We construct a mapping by \emph{aligning} the logged events from each component, as depicted in Figure~\ref{fig:io-mapping}. When doing so, we face two key difficulties. First, implementations may update state memory before or after responding to an input, hence we must ensure we take snapshots to capture both possibilities (\ie case \pC in Figure~\ref{fig:io-mapping}). Second, some parts of a query may not trigger a response, hence we must account for the absence of \textit{write}-like syscalls triggering a snapshot for some queries (\ie cases \pB and \pD).

For all scenarios, we trigger \emph{input} event snapshots after \emph{read}-like syscalls return, and \emph{output} event snapshots just before \emph{write}-like syscalls execute. We take additional snapshots for \emph{output} events before each \textit{read}-like syscall. This enables us to always capture state changes, even if no corresponding \emph{write}-like syscall occurs, or the state change happens after the output is observed.

If there is no output to the final input of a query $q$, then we instruct the learner to pose an additional query with an extra input, $q || i$ for all $i \in I$. We expect there to be a \textit{read} event corresponding to the final input of one of these queries---thus providing us with a snapshot of the state after processing the penultimate input.
If the SUT does not perform such a read, \ie it stops processing new input, as in case \pD, we detect this by intercepting syscalls related to socket closure, and inform the learner that exploration beyond this point is unnecessary.

\subsection{Identifying Candidate State Memory\label{sec:statemem_identify}}

In the first stage of learning, our objective is to identify a set of possible memory locations whose values represent state (Definition 1), \ie $M$. Our analyses aim to find $M$ such that the only locations it contains are values that can be used to discern if we are in a given state, for all possible protocol states that the SUT can be in.

\medskip\noindent\textbf{Snapshot Generation}  To form an initial approximation of $M$, we perform bootstrap queries against the SUT, this produces a set of memory snapshots where all of the memory which tracks state is defined and used (Assumption 3).
Our bootstrap queries take the form, $BF = \{b_0, b_1, ... b_n\}$, where $b_i \in I^+$. The first of these queries $b_0$ is set as the \textit{happy flow} of the protocol. This is the expected normal flow of the protocol execution, which we assume prior knowledge of. For example, in TLS 1.2, $b_0 = (ClientHelloRSA, \allowbreak ClientKeyExchange,\allowbreak  ChangeCipherSpec,\allowbreak  Finished)$. The other queries in $BF$ are specific mutations of this happy flow, automatically constructed with the intention of activating error states, timeout states, and retransmission behaviour. Each of these queries $b_x$ is derived by taking all prefixes $p_x$ of the happy flow $b_0$, where $0 < |p_x| \leq |b_0|$, and for all inputs in $I$, and appending to $p_x$ each $i \in I$ a fixed number of times $T$ such that $b_x = p_x || i^T$.

Every bootstrap query is executed at least twice, so that we have at least two snapshots for each equivalent input sequence, which we require for the next step of our analysis.
When compared to black-box approaches, one might assume that our method requires many more queries in order to facilitate learning. This is not the case. In fact, once we identify a SUT's state memory, we can reuse all of the bootstrap queries for refining our model in subsequent phases.

When possible, we also attempt to alternate functionally equivalent input parameters to the SUT across bootstrap flows. We do this to
maximise the potential number of locations we eliminate due to not holding the same value at equivalent states (Assumption 3). For example, for TLS implementations, we execute some bootstrap flows with different server certificates, which enables us to eliminate memory that would otherwise have to be identified as non-state-defining using our concolic analyser---a comparatively heavyweight procedure.

\begin{figure}[t]
  \centering
    \footnotesize
    \def\svgwidth{0.75\columnwidth}
    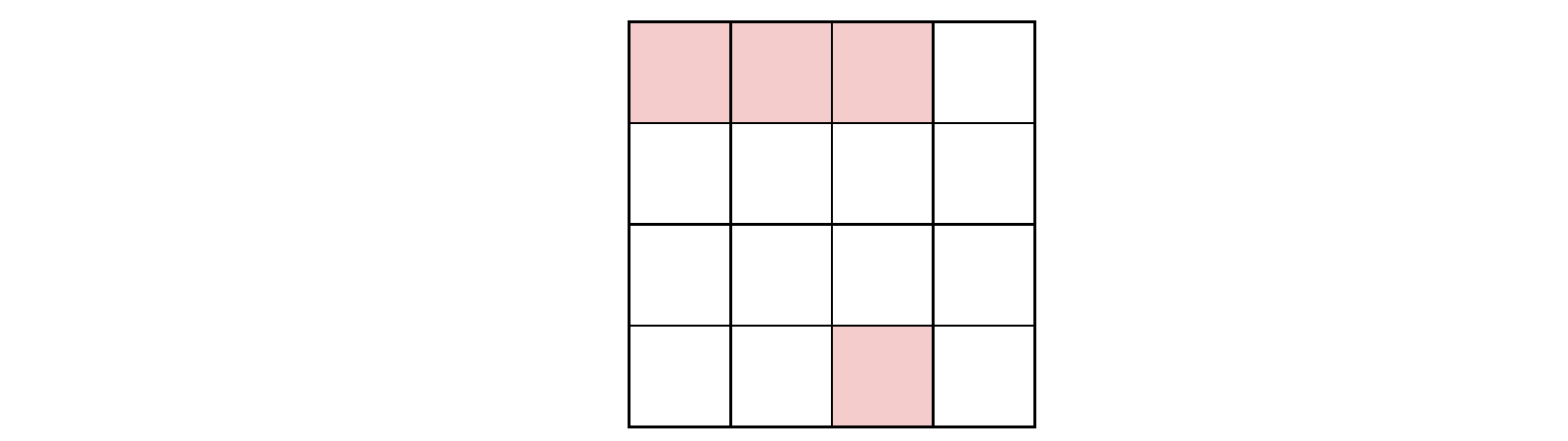
  \caption{Allocation alignment across different executions. Each large block represents the memory layout of a different protocol run; coloured squares represent state-defining locations. Allocation alignment computes a mapping of allocations of a given run (blue squares) to a \emph{base} configuration (red squares). Using this mapping, we can \emph{diff} snapshots that have different configurations, \eg ~\pA and \pC, by first mapping them onto a common configuration, \ie~\pB.
  \label{fig:alloc-align}}
\end{figure}

\medskip\noindent\textbf{Handling Dynamic Memory} To facilitate handling multiple simultaneous sessions, client/server implementations generally allocate state memory on-demand. This presents a challenge when identifying state-defining locations, as logically equivalent allocations may not reside at the same offsets across different protocol executions. To address this non-determinism, we compute a mapping of each execution's allocations to a single \emph{base} configuration. This enables us to analyse all snapshots together with respect to a common configuration, rather than pair-wise. Figure~\ref{fig:alloc-align} visualises our approach. We first construct an allocation log for each execution that records a timestamp, the stack pointer (\ie callee return address), and call parameters, of every call to a memory allocation function (\eg \texttt{malloc} and \texttt{free}). Then, to derive a mapping between two logs, we \emph{align} their allocations/frees, by matching them on their log position, allocation size, and calling context (recorded stack pointer). We choose our \emph{base} configuration, or log, as the largest \emph{happy flow} log, under the assumption that it will contain all allocations related to state-defining locations for any possible session.

\medskip\noindent\textbf{Snapshot Diffing} Following mapping each bootstrap snapshot's allocations onto the \emph{base} log, we \emph{diff} them to obtain our candidate set $M$. Each element $m \in M$ corresponds to the location of a contiguous range of bytes of dynamically allocated memory, which we represent by: an allocation address, an offset relative to the start of the allocation, and a size.

We perform diffing by first grouping all snapshots by their associated I/O sequences. Then, for each group, we locate equivalent allocations across snapshots and identify allocation-offset pairs which refer to byte-sized locations with the same value. We then check that every identified location also contains the same value in every other I/O equivalent snapshot, and is a non-default value in at least one snapshot group (Assumption 3). This gives us a set of candidate state memory locations. We note that as this process is carried out at the \textit{byte} level, we additionally record all bytes that do not abide by this assumption. This enables us to remove any misclassified bytes once we have established the real bounds of individual locations (Section~\ref{sec:typeinf}).

\subsection{Minimising Candidate State Memory}

Given our initial set of candidate state memory locations, we reduce $M$ further by applying the following operations:

\begin{enumerate}
    \item \textbf{Pointer removal:} we eliminate any memory containing pointers to other locations in memory. We do this by excluding values which fall within the address-space of the SUT and its linked libraries.
    \item \textbf{Static allocation elimination:} we remove any full allocations of memory which are assigned a single value in the first snapshot which does not change throughout the course of our bootstrap flows.
    \item \textbf{Static buffer elimination:} we remove any contiguous byte ranges larger than 32 bytes that remain static.
\end{enumerate}

Static allocation and buffer elimination are used to filter locations corresponding to large buffers of non-state-defining memory, for example, in OpenSSL, they eliminate locations storing the TLS certificate.

\subsubsection{Candidate State Memory Type-Inference\label{sec:typeinf}}

The values stored at state-defining locations are often only meaningful when considered as a group, \eg by treating four consecutive bytes as a 32-bit integer. Since we perform snapshot diffing at a byte-level granularity, we may not identify the most-significant bytes of larger integer types if we do not observe their values changing across snapshots. As learning the range of values a location can take is a prerequisite to determine some kinds of termination behaviour, we attempt to monitor locations with respect to their intended types.

To learn each location's bounds, we apply a simple type-inference procedure loosely based upon that proposed by Chen et al.~\cite{chen2018angora}. We perform inference at the same time as we analyse snapshots to handle uncertain state memory locations (Section~\ref{sec:uncertainmem}). During this analysis, we simulate the SUT's execution for a fixed window of instructions, while doing so, we analyse loads and stores from/to our candidate state memory. To perform inference, we log the prefixes used in each access, \eg in \texttt{mov byte ptr [loc\_1], 0}, we record \texttt{byte} for \texttt{loc\_1}. Then, following our main snapshot analysis, we compute the maximal access size for each location and assign each location a corresponding type. To disambiguate overlapped accesses, we determine a location's type based on the minimal non-overlapped range. For example, if we observe that \texttt{loc\_1} and \texttt{loc\_1+2} are both accessed as 4-byte values, we compute \texttt{loc\_1}'s type as two bytes and \texttt{loc\_1+2} as four bytes.
Following type-inference, we update our model and the state classifications maintained by our learner using the new type-bounds discovered.

\subsection{State Merging}
\label{sec:merge}

\begin{figure}[t]
\centering
\scalebox{0.7}{
\footnotesize
\def\svgwidth{0.8\linewidth}
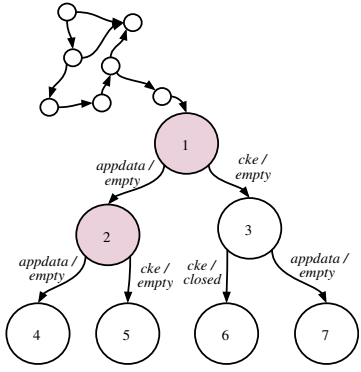
}
\caption{An example scenario where merge conditions for states 1 and 2 are met, when D is set to 1. The state learner would as such taint test any differing memory between the two states to estimate if this memory is state defining.}
\label{fig:merge}
\end{figure}

We consider a unique assignment of candidate state memory a unique state (Definition 2). Hence two states reachable by the same inputs, with equivalent assignments for these locations can be considered equal and merged. However, by applying a trivial equality check, if $M$ is not minimal, \ie it is an over-approximation, this check may not yield the correct outcome. Which, in the worst case could lead to non-termination of learning. We therefore must address this to ensure learning is possible in all cases with respect to Assumption 1, namely that the state machine being learned is finite. The method we present in this section handles this possibility by allowing for states to be merged under the assumption that our candidate set contains superfluous locations.
We attempt to merge states in the model each time all queries in the queue of a given length have been performed. We call this the \textit{merge check}. It operates by identifying pairs of \textit{base} and \textit{merge} states, which can be merged into a single state. These pairs are selected such that two properties hold: \pA the \textit{merge} state must have \textit{I/O equivalence} to the \textit{base}, and \pB the \textit{merge} state should be reachable from the \textit{base}. I/O equivalence signifies all input-output pairs are equal for the two states, to a depth $D$. The intuition is that if I/O differences have not manifested in $D$ steps, they may in fact be the same state, and we should therefore check if the differences in their memory are state relevant. The base-to-merge state reachability must also be possible in $D$ steps (Assumption 4). We enforce this property based on the observation that in security protocol implementations, state duplication is more likely with states along the same \textit{path}. So-called \textit{loopback} transitions often occur where inputs are effectively ignored with respect to the state, however their processing can still result in some changes to memory, resulting in duplication for our memory classified states. Loops in models between multiple states are less common, but will be checked by our learner provided the number of states involved is less than or equal to $D$.

In Figure~\ref{fig:merge}, we present an example of where the mergeability properties hold, with $D = 1$ for states 1 and 2---they are I/O equivalent to depth 1, and state 2 is reachable from state 1. Our merge check determines whether the memory that distinguishes the states is state-defining.
Algorithm~\ref{alg:watchpoints} (see Appendix) shows a sketch of our approach.
In summary, for each location of differing memory, and for each input message (to ensure completeness), we monitor the execution of the SUT. We supply it with input messages to force it into the state we wish to analyse (\ie state 1 or 2 in our example in Figure~\ref{fig:merge}), we then begin to monitor memory reads to the tracked memory location. We follow this by supplying the SUT with each input in turn and perform context snapshots on the reads to the memory location, which we call a \textit{watchpoint hit}.
We then supply each of these snapshots as inputs to our concolic analyser (detailed in the next section), which determines if the SUT uses the value as state-defining memory (Assumption 5). If our analysis identifies any location as state defining, then we do not perform a merge of the two states. Conversely, if all tested locations are reported as not state-defining, then we can merge the \textit{merge} state into the \textit{base} state, and replace the transition with a self loop.

\subsection{Handling Uncertain State Memory\label{sec:uncertainmem}}

Forming our candidate set $M$ by computing the differences between snapshots gives us an over-approximation of all of the locations used discern which state the SUT is in. However, an implementation may not use all of these locations to decide which state it is in for every state. As described in the previous section, when testing if two memory configurations for the same I/O behaviour should be considered equivalent, we must identify if differing values at candidate locations imply that the configurations correspond to different states (\ie the differing locations are state-defining), or if the values they take are not meaningful for the particular state we are checking. For a given state, by analysing how locations actually influence execution we can easily tell state-defining locations from those that are not, as in all cases, non-state-defining locations will have no influence on how state is updated. Since this kind of analysis is expensive, the diffing phase is crucial in minimising the number of candidate locations to analyse---typically reducing the number to tens, rather than thousands.

\subsubsection{Properties Of State-Defining Memory}
\label{sec:propstatedef}

To confirm a given location is state-defining, we attempt to capture execution traces of its location behaving as state-defining memory. We summarise these behaviours below, which we base on our analysis of various implementations:

\begin{enumerate}
    \item \textbf{Control-dependence:} writes to state memory are control-dependent on reads of state memory. To illustrate this, consider the typical case of a state enum flag read forming the basis of a decision for which code should process an incoming message, and the resulting state machine transition defined by a write to state memory.
    \item \textbf{Data-dependence:} non-state memory that is conditionally written to due to dependence on a read from state memory may later influence a write to state memory.
    \item \textbf{State-defining and state-influential locations:} for a particular state, its state-defining memory locations (Definition 2) will always be read from before they are written to (to tell which state one is in), and subsequent writes to state memory will be control- or data-dependent upon the values of those reads (Assumption 5). For state-influencing locations, \eg input buffers, this property will not hold---while the contents of a buffer may \textit{influence} state, it will not directly \textit{define} it, and will be written to before being read.
\end{enumerate}

\subsubsection{Discerning State-Defining Locations\label{sec:taint}}

\begin{figure}[t!]
  \centering
  \footnotesize
  \def\svgwidth{\linewidth}
  \input{st-update-single-column.pdf_tex}
  \caption{OpenSSL client verification bypass. At \pA we check the client certificate signature, at \pB we check if the \texttt{read\_seq} counter is equal to 11, if so, we bypass the check of the result of client certification verification via \pC. At \pD we perform a state update that is control-dependent on \texttt{read\_seq}'s value. \algoName identifies the dependence and dynamically adjusts the learning depth to discover the deep-state change.}
  \label{fig:backdoor-vis}
\end{figure}
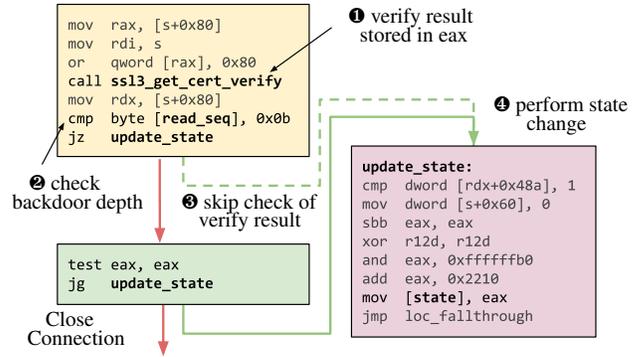

If a location holds more than one value when performing a merge check, we apply an additional analysis to determine if it is state-defining. First, we identify execution paths that are control-dependent on the value stored at the location. Then, we check if any of those paths induce a write to known state memory. If so, we classify the location as state-defining.

We base our analysis on a variation of byte-level taint propagation combined with concolic execution. We apply it in sequence to state snapshots taken on reads of the candidate location $addr$. We start analysis from the instruction $pc$ performing the read that triggered the state snapshot.
Our analysis proceeds by tainting and symbolising $addr$, then tracing forwards until we reach a branch whose condition is tainted by $addr$. If we do not reach a tainted conditional within $W$ instructions, we do not continue analysing the path. At the conditional, we compute two assignments for the value at $addr$---one for each branch destination. We then symbolically explore each of the two paths until we have processed $W$ instructions or we reach a return instruction. If we observe a write to known state memory on either path, we consider $addr$ to be state-defining. Figure~\ref{fig:backdoor-vis} depicts an example location that requires such analysis. The counter \texttt{read\_seq} is used to implement a sneaky backdoor that bypasses client certificate verification in OpenSSL. Our analysis finds that \texttt{read\_seq} taints the branch \texttt{jz update\_state}, and thus, leads to a state change. We therefore classify it as state-defining.

We select an analysis bound of $W=512$ based on the observation that reads from memory involved in comparisons tend to have strong locality to the branches they influence. Since we analyse multiple state snapshots for each $addr$, generated for every possible input in $I$, we reduce the chance for missing locations used in state-defining ways outside of our bound. In practice, we find that our selection of $W$ leads to no false negatives, and note that it can be configured to a larger value to handle problematic implementations.

\section{Implementation}

Our resulting implementation~\cite{github-url} consists of a Java-based learner, a ptrace-based~\cite{ptrace-burrito} execution monitor, and test harnesses in Python and Java. Our concolic analyser is built using Triton~\cite{triton} and IDA Pro~\cite{idapro}. In total, \algoName consists of over 10kloc across three different languages.

\section{Results}
\label{sec:results}

In this section, we present the evaluation of our approach, \algoName, with respect to the research questions outlined in Section~\ref{sec:motivation}. To this end, we test the implementations of two security protocols, TLS and WPA/2. We additionally carry out tests on a intentionally modified TLS implementation, as well as a number of purpose-built protocols.

We address \textbf{RQ1} by demonstrating that our approach makes it possible to learn models with both large input sets and identify the presence of \textit{deep} states in our purpose built protocols and OpenSSL. In doing so, we also show how grey-box observations make it possibly to target and optimise learning (\textbf{RQ2}), thereby avoiding cases of futile state exploration which degrade black-box learning performance. Finally, in order to illustrate the additional insights over I/O that our method affords (\textbf{RQ3}), we discuss our findings regarding the disparity between I/O and memory state classifications of two OpenSSL implementations, as well as an investigation of troublesome state memory in Hostap.

\begin{table*}[t]
\footnotesize
    \centering
    \caption{Model learning results on TLS 1.2 and WPA/2 servers. TLS experiments are repeated with two alphabets, core and extended. Times specified in hours and minutes (hh:mm). Experiments which did not terminate within 3 days are denoted: $\blacksquare$.\label{tbl:results}}
    \begin{tabular}
    {
    P{0.3cm}@{\hspace{0.1cm}} 
    P{1cm}@{\hspace{0.95ex}} 
    P{2.5cm}@{\hspace{0.9ex}} 
    C{1cm} C{1cm} 
    P{1cm}@{\hspace{0.95ex}} 
    P{1cm}@{\hspace{0.95ex}} 
    P{1cm}@{\hspace{0.95ex}} 
    P{1.2cm} 
    P{0.6cm} P{0.5cm}@{\hspace{0.9ex}} 
    P{1.5cm}@{\hspace{0.9ex}} 
    |P{0.7cm}P{0.7cm}} 
    \toprule
        \multirow{2}{0.3cm}{} &
        \multirow{2}{1cm}{\centering \textbf{Protocol}} & \multirow{2}{2.5cm}{\centering \textbf{Implementation}} & \multicolumn{2}{c}{\textbf{Classifying Mem.}} & \multirow{2}{1cm}{\centering \textbf{Mem. States}} & \multirow{2}{1cm}{\centering \textbf{I/O States}} & \multirow{2}{1cm}{\centering \textbf{Total Queries}} &  \multirow{2}{1.2cm}{\centering \textbf{I/O Mem. Queries}} & \multicolumn{2}{c}{ \textbf{Watchpoints}} & \multirow{2}{1.5cm}{\centering \textbf{Total Time}} &
        \multicolumn{2}{c}{\centering \textbf{Black-Box}}  \\
        & & & \textbf{Locations} & \textbf{Allocations} &&&& & \textbf{Queries} & \textbf{Hits} && \textbf{Queries} & \textbf{Time}\\ \midrule
      \parbox[c]{2mm}{\multirow{8}{*}{\rotatebox[origin=c]{90}{\textbf{Core Alpha.}}}}   & TLS 1.2  & RSA-BSAFE-C 4.0.4 & 88  & 14 & 11 & 9 & 128 & 109 & 19 & 4 & 00:06 & 204k & $\blacksquare$ \\
        & TLS 1.2  & Hostap TLS        & 159 & 32 & 11 & 6 & 194 & 160 & 34 & 17 & 00:24 & 971 & 01:48 \\
        & TLS 1.2  & OpenSSL 1.0.1g    & 126 & 16 & 19 & 12 & 488 & 280 & 208 & 116 & 00:18 & 5819 & 00:42\\
        & TLS 1.2  & OpenSSL 1.0.1j    & 125 & 16 & 13 & 9 & 291 & 165 & 126 & 23 & 00:10 & 1826 & 00:13\\
        & TLS 1.2  & OpenSSL 1.1.1g    & 126 & 6 & 6 & 6 & 88 & 86 & 2 & 0 & 00:02 & 175 & 00:02\\
        & TLS 1.2  & GnuTLS 3.3.12     & 172 & 7 & 16 & 7 & 265 & 129 & 36 & 42 & 00:10 & 1353 & 00:21 \\
        & TLS 1.2  & GnuTLS 3.6.14     & 138 & 7 & 18 & 8 & 973 & 452 & 522 & 121 & 00:36 & 3221 & 00:52\\
        \midrule
      \parbox[c]{2mm}{\multirow{5}{*}{\rotatebox[origin=c]{90}{\textbf{Ext. Alpha.}}}}   & TLS 1.2  & RSA-BSAFE-C 4.0.4 & 165 & 7 & 14 & 7 & 918 & 688 & 230 & 635 & 01:08 & 133k & $\blacksquare$  \\
        & TLS 1.2  & OpenSSL 1.1.1g    & 467 & 66 & 11 & 11 & 535 & 524 & 11 & 0 & 01:05 & 1776 & 01:25 \\
        & TLS 1.2  & OpenSSL 1.0.1g    & 554 & 110 & 40 & 20 & 2454 & 1816 & 594 & 609 & 04:21 & 20898 & 21:05 \\
        & TLS 1.2  & GnuTLS 3.3.12     &  185 & 5 & 23 & 19 &  1427 & 1274 & 153 & 128 & 02:05 & 66k & $\blacksquare$ \\
        & TLS 1.2  & GnuTLS 3.6.14     & 157 & 6 & 13 & 11 & 730 & 673 & 57 & 37 & 01:32 & 66k & $\blacksquare$ \\
        \midrule
        & WPA/2 & Hostap 2.8$^\dagger$ & 138 & 3 & 24 & 6 & 1629 & 804 & 825 & 261 & 03:30 & 2127 & 03:30 \\
        & WPA/2 & IWD 1.6 & 135 & 8 & 5 & 5 & 264 & 126 & 138 & 597 & 01:02 & 3000+ & 08:00+ \\
        \bottomrule
    \end{tabular}\\
    {\vspace{1ex}\footnotesize $\dagger$ For Hostap, we stopped both learners at \textapprox 200 minutes, as we found that its state machine is infinite and would prevent both approaches terminating.}
\end{table*}

\begin{figure*}[t]
\centering
\includegraphics[width=\linewidth]{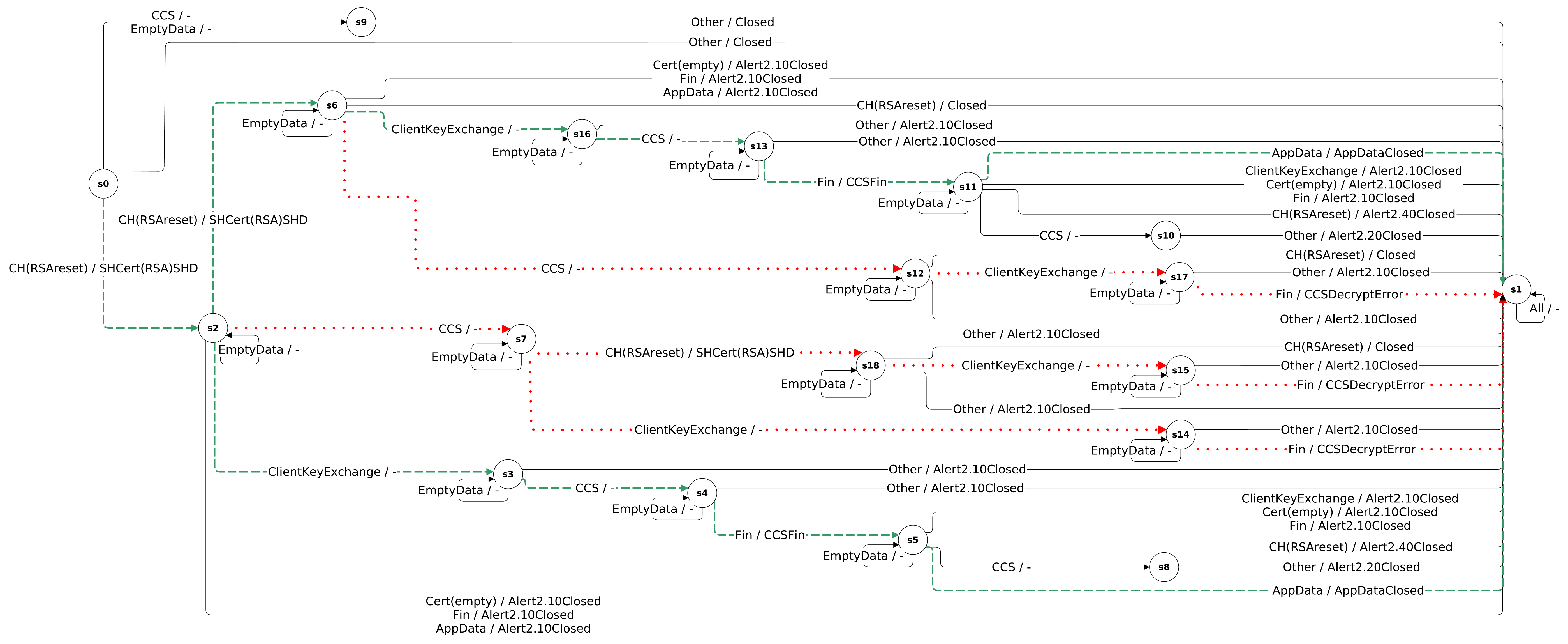}
\caption{Learned model for OpenSSL 1.0.1g}
\label{fig:openssl101g}
\end{figure*}

\subsection{TLS}

We evaluate the interesting examples of TLS 1.2 servers from the work of de Ruiter \etal~\cite{de2015protocol},
the latest versions of these servers, and both the server and client implementations of Hostap's internal TLS library. This spans 4 different code bases, and 7 unique implementations. We learn these models with two separate input sets, one containing the core TLS messages (as in de Ruiter~\etal), and where possible, another with a much larger input set, including client authentication and multiple key exchanges (as provided by the TLS-Attacker test harness~\cite{somorovsky2016systematic}). Table~\ref{tbl:results} lists the time taken to learn all models, as well as the identified \textit{candidate state memory} details, query statistics and number of states identified. Further, we list the number of queries required to learn the same models with state-of-the-art black-box learning, \ie the TTT algorithm~\cite{isberner2014ttt} with the modified W-Method equivalence checking of \cite{de2015protocol} with a conservative equivalence checking depth of 3. For practical reasons, we cut off black-box experiments after 3 days if they show no sign of approaching termination.

\subsubsection{OpenSSL} We tested a number of different versions of OpenSSL, ranging from 2014 to 2020. Of particular interest is the OpenSSL server running version 1.0.1g. A state machine for this implementation was presented at Usenix 2015 \cite{de2015protocol}. It modelled a previously discovered critical vulnerability: specifically, if a \texttt{ChangeCipherSpec} (\texttt{CCS}) message was sent before the \texttt{ClientKeyExchange}, the server would compute the session keys using an empty master secret. Prior to finalising our model, which we depict in Figure~\ref{fig:openssl101g}, we replicated the notably different model presented in \cite{de2015protocol}, by using an identical test harness. As described in Section~\ref{sec:motivation}, this harness sent invalid \texttt{Finished} messages on paths with more than one \texttt{ClientHello}. This results in a model where the dashed green path (s0, s2, s6, s16, s13, s11) and dotted red path (s0, s2, s7, s18, s15) in Figure~\ref{fig:openssl101g} always lead to an \texttt{Alert} and connection closure after receiving the final input. As this was a test harness configuration issue, we sought to use \algoName in order to confirm this fact. In Table~\ref{tbl:opensslmem} (Appendix B), we list the differences in memory between states on alternative paths to handshake completion, and the known happy flow state s4. These differences were identified once learning had terminated. As merge conditions were never met for these states, we carried out a post-analysis stage, where we used \algoName to determine whether the differing memory defined the respective states. We found that it did not.
Despite these states sharing different I/O definitions, when compared to state s4, all were equivalent in terms of state-defining memory, thus indicating that the test harness was responsible for the differences.
This example also illustrates where I/O interpretation by the test harness is misleading. States s15 and s17, which manifest due to the aforementioned early-\texttt{CCS} vulnerability, are also equivalent to the happy flow state s4. The I/O difference in this case is due to the fact that the test harness cannot decrypt the response (hence \texttt{DecryptError}).
We note that this post-learning analysis does not come for free, therefore, we only execute such checks when further model investigation is needed.

When testing client authentication with the extended alphabet, we also discovered inconsistencies in the documentation, which, at worst, have the potential to cause vulnerable misconfigurations. We discuss this further in Appendix D.

\subsubsection{RSA-BSAFE-C} We selected this implementation as it serves to demonstrate that our method is applicable not just to open-source implementations, but also to closed-source. The implementation demonstrates a further advantage of our approach over black-box learning. In particular, we observe that each time the server sends an \texttt{Alert} message, it performs a partial socket shutdown. Specifically, it executes a \texttt{shutdown\string(n, SHUT\_RD\string)} on the connection with the test harness. This means that it no longer accepts input, however, this is not detectable by the test harness. For our approach, this is not a problem; we can observe the \texttt{shutdown} syscall and prevent further inputs from this point. For black-box learning, the socket closure is not detected, and so learning continues exploring beyond the receipt of an \texttt{Alert} message. As shown in Table~\ref{tbl:results}, this leads to many superfluous queries, and, as a result, the learner fails to terminate within 3 days for either alphabet. In comparison, our algorithm is able to learn the same models with 128 queries in 6 minutes for the core TLS functionality, and in \textapprox 1 hour for a more complex model capturing client authentication and alternative key exchanges. Notably, this latter test revealed that repeated \texttt{ClientHello}'s are only permitted when the server is configured with forced client authentication.

\subsubsection{GnuTLS} Our tests on the TLS server implementations in GnuTLS 3.3.12 and 3.6.14 showed substantial changes between the two versions.
In particular, each version required us to hook different syscalls for snapshotting. State count also differed, especially so when testing with the extended alphabet. Analysis of the models revealed slightly different handling of Diffie-Hellman key exchanges, which in the older version resulted in a path of states separate from RSA key exchange paths.

The difference in learning performance between the two approaches, in the case of the extended alphabet, was profound. Black-box learning failed to terminate after 3 days and over 60k queries. We found that this was due to multiple states and inputs warranting empty responses. Consequently, black-box learning exhausted its exploration bound, trying all possible combinations of the troublesome inputs at the affected states. In contrast, as shown in Table~\ref{tbl:results}, \algoName is able to handle such cases much more effectively. This is in part because some groups of inputs are found to result in equivalent snapshots, and when the state equivalence is not immediately evident, our merging strategy quickly finds memory differences are inconsequential.

\subsubsection{Hostap-TLS}

We tested Hostap's internal TLS library as both a client and server.
Although this library is described as experimental, it is used in resource-constrained Wi-Fi devices with limited flash space~\cite{blog-wpasupp-tls} and
by Wi-Fi clients in embedded Linux images created using Buildroot~\cite{buildroot},
\eg motionEyeOS~\cite{motioneyeos}.

Surprisingly, we found that this TLS library always sends TLS alerts in plaintext, which might leak information about the connection.
Further, some frames from the extended alphabet, such as Heartbeat requests or responses were not supported, and sending them resulted in desynchronisation of the TLS connection. We therefore do not include learning results for this implementation with the extended alphabet.

More worrisome, the model showed that against a client, the \texttt{ServerKeyExchange} message can be skipped,
meaning an adversary can instead send \texttt{ServerHelloDone}.
The client will process this message, and then hits an internal error when sending a reply because no RSA key was received to encrypt the premaster secret.
When Ephemeral Diffie-Hellman is used instead, the client calculates $g^{cs}$ as the premaster secret with $s$ equal to zero if no \texttt{ServerKeyExchange} message was received.
Because the exponent is zero, the default math library of Hostap returns an internal error, causing the connection to close, meaning an adversary cannot abuse this to attack clients.
Nevertheless, this does illustrate that the state machine of the client does not properly validate the order of messages.

\subsection{WPA/2's 4-Way Handshake}

We also tested WPA/2's 4-way handshake which is used to authenticate with Wi-Fi networks.
There are two open source implementations of this handshake on Linux, namely the ones in IWD~(iNet Wireless Daemon) and Hostap, and we test both.

To learn the state machine, we start with an input alphabet of size 4 that only contains messages which occur in normal handshake executions.
Our test harness automatically assigns sensible values to all of the fields of these handshake messages.
To produce larger input sets, we also tried non-standard values for certain fields.
For example, for the replay counter, we tried the same value consecutively, set it equal to zero, and other variations. To this end, we created two extended alphabets: one of size 15 and another one of size 40.

To also detect key reinstallation bugs~\cite{vanhoef2017keyreinstallation}, we let the SUT send an encrypted dataframe after completing the handshake.
In the inferred model, resulting dataframes encrypted using a nonce equal to one are represented using \verb|AES_DATA_1|,
while all other dataframes are represented using \verb|AES_DATA_n|.
A key reinstallation bug, or a variant thereof, can now be automatically detected by scanning for paths in the inferred model that contain multiple occurrences of \verb|AES_DATA_1|.

One obstacle that we encountered is handling retransmitted handshake messages that were triggered due to timeouts.
Because this issue is orthogonal to evaluating our memory-based state inference method, we disabled retransmissions, and leave the handling of retransmissions as future work.

\subsubsection{IWD}

We found that the state machine of IWD does not enter a new state after receiving message 4 of the handshake.
This means an adversary can replay message 4, after which it will be processed by IWD, triggering a key reinstallation.
Note that this is not a default key reinstallation, \ie
we confirmed that existing key reinstallation tools cannot detect it~\cite{krack-scripts}.
The discovered reinstallation can be abused to replay, decrypt, and possibly forge data frames~\cite{vanhoef2017keyreinstallation}.
We reported this vulnerability and it has been patched and assigned CVE-2020-NNNNN~\cite{iwd-cve}.
When running our tool on the patched version of IWD, the state machine enters a new state after receiving message 4, which confirms that the vulnerability has been patched.

With black-box testing, the key reinstallation is only found when using a small input alphabet \emph{and} when the test harness always sends handshake messages with a replay counter equal to the last one used by the AP.
If the harness instead increments the replay counter after sending each handshake message, the key reinstallation can only be discovered when using an extended alphabet.
However, in that case black-box learning takes much longer: after 3 hours the learner creates a hypothesis model that includes the vulnerability, but it is unable to verify this hypothesis within 8 hours (at which point we terminated the learner).
This shows that our method handles larger input alphabets more efficiently, especially when queries are slow, resulting in more accurate models and increasing the chance of finding bugs.

\subsubsection{Hostapd}

One obstacle with Hostapd is how the last used replay counter in transmitted handshake messages is saved:
our learner initially includes this counter in its candidate state memory set.
When trying to merge states with different replay counter values, our concolic analysis determines this value to be state defining. Indeed, Hostapd checks whether incoming frames include a particular value. However, with each loop of the state machine, the expected value changes, and hence the learner prevents the merge. This results in a violation of Assumption~1, meaning learning will not terminate.
Like with other violations of assumptions, we addressed this by including a time bound on the learning.
Because of this we also do not explore a bigger alphabet.
For a fair comparison, we used the same time bound in our method and in black-box learning. The models which were produced with both approaches within \textapprox 200 minutes were equivalent.
Under these conditions, we note that both resulting state machines followed the standard and contained no surprising transitions.

\subsection{Example Protocols}

In order to highlight the capability of \algoName to identify states deep in a protocol, we test a minimally modified version of OpenSSL, in addition to a variety of example protocols. We note that states in protocols beyond typical learning bounds are not only theoretical---a state machine flaw in a WPA/2 router \cite{stone2018extending} found a cipher downgrade was possible after 3 failed initiation attempts (which would have been missed with configured bounds less than 3, as in \cite{de2015protocol}).

Our modification of OpenSSL (version 1.0.1j), consists of a 4 line addition (Appendix~\ref{sec:backdoor-apdx}, Listing~\ref{lst:backdoor}), which hijacks existing state data to implement a simple client authentication bypass, activated by $n$ unexpected messages (see Figure~\ref{fig:backdoor-vis} of Section~\ref{sec:taint}). To learn this model, we used an extended alphabet of 21 messages, including core TLS functionality, client certificates, and various data frames. With $n=5$, black-box learning configured with a bound equal to $n$, fails to identify the state after 3 days of learning and \textapprox100k queries (and showed no signs of doing so soon). \algoName, on the other hand, identifies the main backdoor-activation state in under 2 hours and \textapprox1.2k queries and the second activation variant (with two preceding \texttt{ClientHellos}, as opposed to the usual one) in under 3 hours and \textapprox2.5k queries. Doubling the depth $n$ results in 3.2k and 4k queries respectively---an approximately linear increase. Black-box learning predictably failed to locate the even deeper state due to its exponential blow up in this particular scenario.

Our motivating example protocol of Section~\ref{sec:motivation}, Figure~\ref{fig:blackbox-fail} also showed a dramatic advantage over black-box learning, even with a simple model and small input alphabet. We implemented this protocol similarly to a real protocol, with state maintained between an enum and counting integer combined. With black-box learning, this model took 360k queries and 20 hours to identify the auth-bypass but failed to terminate its analysis after 4 days. \algoName identified the backdoor in 149 queries in just over minute, with the concolic analyser taking up 20 seconds of that time.

\section{Discussion}

We have shown that relaxing the restriction of pure black-box analysis improves practical model learning: we can learn new types of states and gain more insights, all in less time. While we have shown our assumptions make a justifiable trade-off between practicality and completeness, we acknowledge that they may not hold for all protocol implementations. Thus, in this section, we discuss possible limitations.

If the machine is not finite (Assumption 1), then, like black-box methods, our approach will fail to terminate. Likewise, with respect to Assumption 4, if loops within a state machine exist beyond the configured depth bound, and the memory across these states is always changing (\eg a counter), then \algoName will interminably duplicate states. In both of these cases, like black-box methods, we provide a means to bound exploration. We choose to bound by a configurable time, but this can also be done by exploration depth. In the worst case, the resulting models will be no less representative than those learned with black-box methods.
False negatives due to missed candidate state memory and concolic analysis also have the potential to be problematic. As a consequence of Assumption 3, we may miss candidate locations if they do not change during any bootstrap flow. An example of this kind of memory is one allocated only after entering an error state.
Fortunately, our results and analysis of implementation code indicate that this assumption holds for most protocols.

An inherent requirement of grey-box analysis is the ability to introspect a program's execution state. Whilst we ensure this is performed as non-invasively as possible, it does limit the types of SUT which can be tested. For example, our method would not be appropriate for protocols running on certain types of embedded device, \eg EMV cards. However, for devices with sufficient debug access, \eg JTAG, \algoName could be used to learn implementations for which black-box learning proves to be too slow.

Finally, although \algoName is designed to work with both closed and open-source implementations, it is often the case---as with most of our tested protocols---that we do have access to source code. If this is the case, our tool is able to map uses of state memory to source code locations---which substantially aids in the analysis of inferred models.

\section{Future Work}

One key contribution of \algoName is its ability to identify the structures within a program which are used to maintain its runtime state. Further, locations within the program which manipulate state memory can also be easily identified. Recent works in the domain of fuzzing (protocols~\cite{chen2019exploring,profuzzbench} and more general software~\cite{aschermann2020ijon,invariantfuzz}) have sought to combine program state coverage with traditional CFG block coverage for fuzzing feedback. These efforts have hitherto required a manual specification of state changing locations---something we believe can be aided by the approaches of \algoName.

\section{Conclusion}

Black-box testing is fundamentally limited because it can only reason about how a SUT interacts with the outside world. On the one hand, this means that a large number of queries are required to learn state machines even for simple protocols, and on the other, because we have no insight into \emph{how} the SUT processes inputs, we cannot optimise queries to trigger or avoid certain behaviours.

In contrast, our grey-box approach, \algoName, overcomes many of these issues. It is able to efficiently handle larger input spaces and protocols that trigger state changes at non-trivial depths, resulting in more accurate and insightful models, that can be learned in substantially fewer queries. As testament to this, we were able to discover a new, high impact vulnerability in IWD, and numerous concerning deviations from the standards in other implementations.

\bibliographystyle{plain}
\bibliography{sample}

\clearpage
\appendix

\section{Algorithm to Generate Watchpoint Hit Snapshots}
\label{sec:algorithm-apdx}
\begin{algorithm}[h!]
\SetAlgoLined
\KwResult{list of $watchpointHitExecutionDumps$}
\KwIn{$(addr, size)$ of memory to test, prefix query $p$ to arrive at $mergeState$, the $SUT$}
 \ForEach{$(input, st) \in reachableFrm(mergeState, 1)$}{
    \If{$outputFor(input, st) \in disabled$}{$continue;$}
    $initialiseWatchpointMonitor(SUT, addr, size)$\;
    $executeQuery(SUT, p || input)$;
 }
 \caption{Generation of watchpoint hit snapshots.}
 \label{alg:watchpoints}
\end{algorithm}

\section{OpenSSL State Memory Analysis}
\begin{table}[h!]
    \caption{The sets of differing memory for states on suspected and confirmed alternate paths to successful handshake completion, when compared to the legitimate \textit{happy flow} state 4 (s4) memory.}
    \centering
    \begin{tabular}{r c l}
    \toprule
     State IDs    & & Addresses of differing memory to s4 \\ \midrule
     s14 & \ldelim\{{2}{2mm}[] \ldelim\{{11}{2mm}[] &\{0x555555a162a0, 0x4b0, 0x13\} \\
     & &\{0x555555a162a0, 0x4b0, 0x120\} \\
                     && \{0x555555a162a0,  0x4b0, 0x0\} \\
                     && \{0x555555a0e6b0,  0x160, 0x0\} \\
                     && \{0x555555a0e6b0,  0x160, 0x10\} \\
     s15, s17, s13   && \{0x555555a0e6b0,  0x160, 0x44\} \\
                     && \{0x555555a0e6b0,  0x160, 0x68\} \\
                     && \{0x555555a0e6b0,  0x160, 0x6c\} \\
                     && \{0x555555a0e6b0,  0x160, 0xc0\} \\
                     && \{0x555555a0e6b0,  0x160, 0xc8\} \\
                     && \{0x555555a0e6b0,  0x160, 0xd2\} \\ \bottomrule
\end{tabular} \label{tbl:opensslmem}
\end{table}

Using the learning output of \algoName, we carried out an investigation into what the differing memory in Table~\ref{tbl:opensslmem} actually constituted. We were able to do this by referring to the logs produced by the tool, which print the source code details of any watchpoint hits made at reads of the tested memory. All considered states shared the two differing addresses listed at the top of the table. The first of these represented a counter referred to as the \texttt{sequence number}. This value increments for each message received after the \texttt{ChangeCipherSpec} message and is used for constructing MACs. The second address in the table refers to the frame type of the just received message. This value is determined as not state defining because from any given state it is always written to before being read, \ie it may be state influencing, but not defining (ref. \ref{sec:propstatedef}). The third address in the table pertains to a flag which determines if the client has attempted more than one \texttt{ClientHello} message. Only prior to the receipt of a \texttt{ClientKeyExchange} is this flag ever read (and therefore used to define the state). Consequently, it is not considered state defining for any of the states listed in the table. The remaining memory refers to arbitrary certificate data which we also found was not state defining. In particular, we were unable to detect any reads of the memory from the states considered.

\section{OpenSSL Client Authentication Bypass Backdoor}
\label{sec:backdoor-apdx}
\begin{code}
\captionof{listing}{A four line modification to OpenSSL 1.0.1j, which introduces a client authentication bypass at a configurable depth (\texttt{BACKDOOR\_DEPTH}). In the first part, we hijack a counter, the \texttt{read\_sequence} counter, and conditionally increment it each time an unexpected data frame is read whilst waiting on the ClientCertificateVerify signature.  In the second part, if this counter reaches the specified depth value, invalid ClientCertificateVerify signatures are ignored, and the handshake proceeds to the ChangeCipherSpec state. After this point, the \texttt{read\_sequence} is automatically reset by the implementation for use in later MAC calculations.}
\label{lst:backdoor}
\begin{minted}[mathescape,
               %linenos,
               numbersep=5pt,
               gobble=2,
               frame=lines,
               escapeinside=||,
               framesep=2mm]{diff}
|\setcounter{FancyVerbLine}{138}|
@@ /ssl/s3_pkt.c:139 ssl3_read_n()
 	SSL3_BUFFER *rb;
+++ // Backdoor pt 1: Set-up
+++ if(s->state == SSL3_ST_SR_CERT_VRFY_A)
+++     ++(*s->s3->read_sequence);

 	if (n <= 0) return n;
 	rb = &(s->s3->rbuf);
|\setcounter{FancyVerbLine}{682}|
@@ /ssl/s3_srvr.c:683  ssl3_accept()
 	ret=ssl3_get_cert_verify(s);
--- if (ret <= 0) goto end;
+
+++ // Backdoor pt 2: Activation
+++ if(*s->s3->read_sequence != BACKDOOR_DEPTH
+++     && ret <= 0) {
+++     goto end;
+++ } else {
      ...
      s->state=SSL3_ST_SR_FINISHED_A;
      s->init_num=0;
+++ }
\end{minted}
\end{code}

\section{Misleading OpenSSL Documentation}

When testing the built-in server of OpenSSL, we found that configuring it to enforce client authentication is rather error-prone. In particular, after an initial reading of the documentation for how to do so, we came across the option below:
\begin{verbatim}
-verify int, -Verify int
    The verify depth to use. This specifies the
    maximum length of the client certificate
    chain and makes the server request a
    certificate from the client. With the
    -verify  option a certificate is requested
    but the client does not have to send one,
    with the -Verify option the client must
    supply a certificate or an error occurs.
\end{verbatim}
Following this, we used the \texttt{-Verify} option when starting the built-in server. Surprisingly though, our learned models indicated that the server accepted invalid client certificates. At first glance, this looked like a severe vulnerability, however, upon closer inspection of the documentation we found an extra option was required:
\begin{verbatim}
-verify_return_error
    Verification errors normally just print a
    message but allow the connection to
    continue, for debugging purposes. If this
    option is used, then verification errors
    close the connection.
\end{verbatim}
Our online investigation revealed we were not alone in our confusion. A recent Github issue\footnote{\url{https://github.com/openssl/openssl/issues/8079}} describes the same misleading configuration, and resulted in a patch to the OpenSSL client implementation\footnote{\url{https://github.com/openssl/openssl/pull/8080}}. Unfortunately, a similar patch was not written for the server implementation, resulting in a confusing disparity between the command-line options for the client and server. We have reported this issue to the OpenSSL developers.

\end{document}

%% file: exampleProtocolv2.pdf_tex
\begingroup%
  \makeatletter%
  \providecommand\color[2][]{%
    \errmessage{(Inkscape) Color is used for the text in Inkscape, but the package 'color.sty' is not loaded}%
    \renewcommand\color[2][]{}%
  }%
  \providecommand\transparent[1]{%
    \errmessage{(Inkscape) Transparency is used (non-zero) for the text in Inkscape, but the package 'transparent.sty' is not loaded}%
    \renewcommand\transparent[1]{}%
  }%
  \providecommand\rotatebox[2]{#2}%
  \newcommand*\fsize{\dimexpr\f@size pt\relax}%
  \newcommand*\lineheight[1]{\fontsize{\fsize}{#1\fsize}\selectfont}%
  \ifx\svgwidth\undefined%
    \setlength{\unitlength}{240bp}%
    \ifx\svgscale\undefined%
      \relax%
    \else%
      \setlength{\unitlength}{\unitlength * \real{\svgscale}}%
    \fi%
  \else%
    \setlength{\unitlength}{\svgwidth}%
  \fi%
  \global\let\svgwidth\undefined%
  \global\let\svgscale\undefined%
  \makeatother%
  \begin{picture}(1,0.79166665)%
    \lineheight{1}%
    \setlength\tabcolsep{0pt}%
    \put(0,0){\includegraphics[width=\unitlength,page=1]{exampleProtocolv2.pdf}}%
    \put(0.68887366,0.6718271){\color[rgb]{0,0,0}\makebox(0,0)[lt]{\lineheight{1.25}\smash{\begin{tabular}[t]{l}1\end{tabular}}}}%
    \put(0,0){\includegraphics[width=\unitlength,page=2]{exampleProtocolv2.pdf}}%
    \put(0.34716843,0.7103849){\color[rgb]{0,0,0}\makebox(0,0)[lt]{\lineheight{1.25}\smash{\begin{tabular}[t]{l}0\end{tabular}}}}%
    \put(0,0){\includegraphics[width=\unitlength,page=3]{exampleProtocolv2.pdf}}%
    \put(0.23033035,0.16783626){\color[rgb]{0,0,0}\makebox(0,0)[lt]{\lineheight{1.25}\smash{\begin{tabular}[t]{l}2\end{tabular}}}}%
    \put(0,0){\includegraphics[width=\unitlength,page=4]{exampleProtocolv2.pdf}}%
    \put(0.75170543,0.17653042){\color[rgb]{0,0,0}\makebox(0,0)[lt]{\lineheight{1.25}\smash{\begin{tabular}[t]{l}3\end{tabular}}}}%
    \put(0,0){\includegraphics[width=\unitlength,page=5]{exampleProtocolv2.pdf}}%
    \put(0.15700386,0.70151324){\color[rgb]{0,0,0}\makebox(0,0)[lt]{\lineheight{1.25}\smash{\begin{tabular}[t]{l}\textit{a}\textit{u}\textit{t}\textit{h}\textit{ }\textit{/}\textit{ }\textit{e}\textit{r}\textit{r}\textit{o}\textit{r}\end{tabular}}}}%
    \put(0.15700386,0.67026347){\color[rgb]{0,0,0}\makebox(0,0)[lt]{\lineheight{1.25}\smash{\begin{tabular}[t]{l}\textit{d}\textit{a}\textit{t}\textit{a}\textit{ }\textit{/}\textit{ }\textit{e}\textit{r}\textit{r}\textit{o}\textit{r}\end{tabular}}}}%
    \put(0.15354319,0.63901369){\color[rgb]{0,0,0}\makebox(0,0)[lt]{\lineheight{1.25}\smash{\begin{tabular}[t]{l}\textit{c}\textit{l}\textit{o}\textit{s}\textit{e}\textit{ }\textit{/}\textit{ }\textit{e}\textit{r}\textit{r}\textit{o}\textit{r}\end{tabular}}}}%
    \put(0,0){\includegraphics[width=\unitlength,page=6]{exampleProtocolv2.pdf}}%
    \put(0.49237255,0.73954201){\color[rgb]{0,0,0}\makebox(0,0)[lt]{\lineheight{1.25}\smash{\begin{tabular}[t]{l}\textit{i}\textit{n}\textit{i}\textit{t}\textit{ }\textit{/}\textit{ }\textit{a}\textit{c}\textit{k}\end{tabular}}}}%
    \put(0,0){\includegraphics[width=\unitlength,page=7]{exampleProtocolv2.pdf}}%
    \put(0.48460171,0.06089393){\color[rgb]{0,0,0}\makebox(0,0)[lt]{\lineheight{1.25}\smash{\begin{tabular}[t]{l}-\end{tabular}}}}%
    \put(0,0){\includegraphics[width=\unitlength,page=8]{exampleProtocolv2.pdf}}%
    \put(0.33690882,0.47511081){\color[rgb]{0.59607843,0,0}\makebox(0,0)[lt]{\lineheight{1.25}\smash{\begin{tabular}[t]{l}\textit{i}\textit{n}\textit{i}\textit{t}\textit{ }\textit{/}\textit{ }\textit{a}\textit{c}\textit{k}\textit{ }\end{tabular}}}}%
    \put(0.3679456,0.44073606){\color[rgb]{0.59607843,0,0}\makebox(0,0)[lt]{\lineheight{1.25}\smash{\begin{tabular}[t]{l}\textit{x}\textit{1}\textit{1}\end{tabular}}}}%
    \put(0,0){\includegraphics[width=\unitlength,page=9]{exampleProtocolv2.pdf}}%
    \put(0.87521571,0.22448255){\color[rgb]{0,0,0}\makebox(0,0)[lt]{\lineheight{1.25}\smash{\begin{tabular}[t]{l}\textit{a}\textit{u}\textit{t}\textit{h}\textit{ }\textit{/}\textit{ }\textit{e}\textit{r}\textit{r}\textit{o}\textit{r}\end{tabular}}}}%
    \put(0.87521571,0.19323277){\color[rgb]{0,0,0}\makebox(0,0)[lt]{\lineheight{1.25}\smash{\begin{tabular}[t]{l}\textit{d}\textit{a}\textit{t}\textit{a}\textit{ }\textit{/}\textit{ }\textit{e}\textit{r}\textit{r}\textit{o}\textit{r}\end{tabular}}}}%
    \put(0.88076987,0.16198299){\color[rgb]{0,0,0}\makebox(0,0)[lt]{\lineheight{1.25}\smash{\begin{tabular}[t]{l}\textit{i}\textit{n}\textit{i}\textit{t}\textit{ }\textit{/}\textit{ }\textit{e}\textit{r}\textit{r}\textit{o}\textit{r}\end{tabular}}}}%
    \put(0,0){\includegraphics[width=\unitlength,page=10]{exampleProtocolv2.pdf}}%
    \put(0.78663781,0.47936095){\color[rgb]{0,0,0}\makebox(0,0)[lt]{\lineheight{1.25}\smash{\begin{tabular}[t]{l}\textit{d}\textit{a}\textit{t}\textit{a}\textit{ }\textit{/}\textit{ }\textit{e}\textit{r}\textit{r}\textit{o}\textit{r}\end{tabular}}}}%
    \put(0.78317714,0.44811118){\color[rgb]{0,0,0}\makebox(0,0)[lt]{\lineheight{1.25}\smash{\begin{tabular}[t]{l}\textit{c}\textit{l}\textit{o}\textit{s}\textit{e}\textit{ }\textit{/}\textit{ }\textit{e}\textit{r}\textit{r}\textit{o}\textit{r}\end{tabular}}}}%
    \put(0,0){\includegraphics[width=\unitlength,page=11]{exampleProtocolv2.pdf}}%
    \put(0.53365965,0.00633106){\color[rgb]{0,0,0}\makebox(0,0)[lt]{\lineheight{1.25}\smash{\begin{tabular}[t]{l}\textit{a}\textit{l}\textit{l}\textit{ }\textit{/}\textit{ }\textit{-}\end{tabular}}}}%
    \put(0,0){\includegraphics[width=\unitlength,page=12]{exampleProtocolv2.pdf}}%
    \put(0.43757254,0.15691337){\color[rgb]{0,0,0}\makebox(0,0)[lt]{\lineheight{1.25}\smash{\begin{tabular}[t]{l}\textit{i}\textit{n}\textit{i}\textit{t}\textit{ }\textit{/}\textit{ }\textit{e}\textit{r}\textit{r}\textit{o}\textit{r}\end{tabular}}}}%
    \put(0,0){\includegraphics[width=\unitlength,page=13]{exampleProtocolv2.pdf}}%
    \put(0.19145113,0.06485846){\color[rgb]{0,0,0}\makebox(0,0)[lt]{\lineheight{1.25}\smash{\begin{tabular}[t]{l}\textit{c}\textit{l}\textit{o}\textit{s}\textit{e}\textit{ }\textit{/}\textit{ }\textit{c}\textit{l}\textit{o}\textit{s}\textit{e}\end{tabular}}}}%
    \put(0,0){\includegraphics[width=\unitlength,page=14]{exampleProtocolv2.pdf}}%
    \put(0.68053895,0.07438108){\color[rgb]{0,0,0}\makebox(0,0)[lt]{\lineheight{1.25}\smash{\begin{tabular}[t]{l}\textit{c}\textit{l}\textit{o}\textit{s}\textit{e}\textit{ }\textit{/}\textit{ }\textit{c}\textit{l}\textit{o}\textit{s}\textit{e}\end{tabular}}}}%
    \put(0,0){\includegraphics[width=\unitlength,page=15]{exampleProtocolv2.pdf}}%
    \put(0.01776005,0.16080597){\color[rgb]{0,0,0}\makebox(0,0)[lt]{\lineheight{1.25}\smash{\begin{tabular}[t]{l}\textit{a}\textit{u}\textit{t}\textit{h}\textit{ }\textit{/}\textit{ }\textit{a}\textit{c}\textit{k}\end{tabular}}}}%
    \put(0.01776005,0.12955619){\color[rgb]{0,0,0}\makebox(0,0)[lt]{\lineheight{1.25}\smash{\begin{tabular}[t]{l}\textit{d}\textit{a}\textit{t}\textit{a}\textit{ }\textit{/}\textit{ }\textit{a}\textit{c}\textit{k}\end{tabular}}}}%
    \put(0,0){\includegraphics[width=\unitlength,page=16]{exampleProtocolv2.pdf}}%
    \put(0.13866875,0.46677345){\color[rgb]{0.21960784,0.4627451,0.11372549}\makebox(0,0)[lt]{\lineheight{1.25}\smash{\begin{tabular}[t]{l}\textit{a}\textit{u}\textit{t}\textit{h}\textit{ }\textit{/}\textit{ }\textit{a}\textit{c}\textit{k}\end{tabular}}}}%
    \put(0,0){\includegraphics[width=\unitlength,page=17]{exampleProtocolv2.pdf}}%
    \put(0.59850344,0.46533955){\color[rgb]{0,0,0}\makebox(0,0)[lt]{\lineheight{1.25}\smash{\begin{tabular}[t]{l}\textit{a}\textit{u}\textit{t}\textit{h}\textit{ }\textit{/}\textit{ }\textit{e}\textit{r}\textit{r}\textit{o}\textit{r}\end{tabular}}}}%
    \put(0.59850344,0.43408978){\color[rgb]{0,0,0}\makebox(0,0)[lt]{\lineheight{1.25}\smash{\begin{tabular}[t]{l}\textit{d}\textit{a}\textit{t}\textit{a}\textit{ }\textit{/}\textit{ }\textit{e}\textit{r}\textit{r}\textit{o}\textit{r}\end{tabular}}}}%
    \put(0.59504278,0.40284){\color[rgb]{0,0,0}\makebox(0,0)[lt]{\lineheight{1.25}\smash{\begin{tabular}[t]{l}\textit{c}\textit{l}\textit{o}\textit{s}\textit{e}\textit{ }\textit{/}\textit{ }\textit{e}\textit{r}\textit{r}\textit{o}\textit{r}\end{tabular}}}}%
    \put(0,0){\includegraphics[width=\unitlength,page=18]{exampleProtocolv2.pdf}}%
  \end{picture}%
\endgroup%

%% file: gb-state-learner.pdf_tex
\begingroup%
  \makeatletter%
  \providecommand\color[2][]{%
    \errmessage{(Inkscape) Color is used for the text in Inkscape, but the package 'color.sty' is not loaded}%
    \renewcommand\color[2][]{}%
  }%
  \providecommand\transparent[1]{%
    \errmessage{(Inkscape) Transparency is used (non-zero) for the text in Inkscape, but the package 'transparent.sty' is not loaded}%
    \renewcommand\transparent[1]{}%
  }%
  \providecommand\rotatebox[2]{#2}%
  \newcommand*\fsize{\dimexpr\f@size pt\relax}%
  \newcommand*\lineheight[1]{\fontsize{\fsize}{#1\fsize}\selectfont}%
  \ifx\svgwidth\undefined%
    \setlength{\unitlength}{582bp}%
    \ifx\svgscale\undefined%
      \relax%
    \else%
      \setlength{\unitlength}{\unitlength * \real{\svgscale}}%
    \fi%
  \else%
    \setlength{\unitlength}{\svgwidth}%
  \fi%
  \global\let\svgwidth\undefined%
  \global\let\svgscale\undefined%
  \makeatother%
  \begin{picture}(1,0.24054983)%
    \lineheight{1}%
    \setlength\tabcolsep{0pt}%
    \put(0,0){\includegraphics[width=\unitlength,page=1]{gb-state-learner.pdf}}%
    \put(0.07248876,0.14024156){\color[rgb]{0,0,0}\makebox(0,0)[lt]{\lineheight{1.25}\smash{\begin{tabular}[t]{l}Learner\end{tabular}}}}%
    \put(0,0){\includegraphics[width=\unitlength,page=2]{gb-state-learner.pdf}}%
    \put(0.21740661,0.02624107){\color[rgb]{0,0,0}\makebox(0,0)[lt]{\lineheight{1.25}\smash{\begin{tabular}[t]{l}Concolic analyser\end{tabular}}}}%
    \put(0,0){\includegraphics[width=\unitlength,page=3]{gb-state-learner.pdf}}%
    \put(0.43989442,0.20281415){\color[rgb]{0,0,0}\makebox(0,0)[lt]{\lineheight{1.25}\smash{\begin{tabular}[t]{l}Test harness\end{tabular}}}}%
    \put(0,0){\includegraphics[width=\unitlength,page=4]{gb-state-learner.pdf}}%
    \put(0.66244725,0.14024156){\color[rgb]{0,0,0}\makebox(0,0)[lt]{\lineheight{1.25}\smash{\begin{tabular}[t]{l}SUT\end{tabular}}}}%
    \put(0,0){\includegraphics[width=\unitlength,page=5]{gb-state-learner.pdf}}%
    \put(0.42934431,0.10736896){\color[rgb]{0,0,0}\makebox(0,0)[lt]{\lineheight{1.25}\smash{\begin{tabular}[t]{l}API interceptor\end{tabular}}}}%
    \put(0,0){\includegraphics[width=\unitlength,page=6]{gb-state-learner.pdf}}%
    \put(0.4152512,0.05214611){\color[rgb]{0,0,0}\makebox(0,0)[lt]{\lineheight{1.25}\smash{\begin{tabular}[t]{l}Watchpoint handler\end{tabular}}}}%
    \put(0,0){\includegraphics[width=\unitlength,page=7]{gb-state-learner.pdf}}%
    \put(0.63713102,0.20241166){\color[rgb]{0,0,0}\makebox(0,0)[lt]{\lineheight{1.25}\smash{\begin{tabular}[t]{l}I/O channel\end{tabular}}}}%
    \put(0,0){\includegraphics[width=\unitlength,page=8]{gb-state-learner.pdf}}%
    \put(0.62568356,0.07975585){\color[rgb]{0,0,0}\makebox(0,0)[lt]{\lineheight{1.25}\smash{\begin{tabular}[t]{l}Debug channel\end{tabular}}}}%
    \put(0,0){\includegraphics[width=\unitlength,page=9]{gb-state-learner.pdf}}%
    \put(0.18377881,0.20241166){\color[rgb]{0,0,0}\makebox(0,0)[lt]{\lineheight{1.25}\smash{\begin{tabular}[t]{l}Output queries\end{tabular}}}}%
    \put(0,0){\includegraphics[width=\unitlength,page=10]{gb-state-learner.pdf}}%
    \put(0.04254516,0.03655197){\color[rgb]{0,0,0}\makebox(0,0)[lt]{\lineheight{1.25}\smash{\begin{tabular}[t]{l}Type \& memory\end{tabular}}}}%
    \put(0.05053296,0.01593346){\color[rgb]{0,0,0}\makebox(0,0)[lt]{\lineheight{1.25}\smash{\begin{tabular}[t]{l}classifications\end{tabular}}}}%
    \put(0,0){\includegraphics[width=\unitlength,page=11]{gb-state-learner.pdf}}%
    \put(0.79678383,0.15979296){\color[rgb]{0,0,0}\makebox(0,0)[lt]{\lineheight{1.25}\smash{\begin{tabular}[t]{l}Black-box component\end{tabular}}}}%
    \put(0.80251176,0.12886519){\color[rgb]{0,0,0}\makebox(0,0)[lt]{\lineheight{1.25}\smash{\begin{tabular}[t]{l}Grey-box component\end{tabular}}}}%
    \put(0.8206264,0.09793741){\color[rgb]{0,0,0}\makebox(0,0)[lt]{\lineheight{1.25}\smash{\begin{tabular}[t]{l}Execution monitor\end{tabular}}}}%
    \put(0.86401777,0.06700964){\color[rgb]{0,0,0}\makebox(0,0)[lt]{\lineheight{1.25}\smash{\begin{tabular}[t]{l}Input/output\end{tabular}}}}%
    \put(0,0){\includegraphics[width=\unitlength,page=12]{gb-state-learner.pdf}}%
    \put(0.1339784,0.07975585){\color[rgb]{0,0,0}\makebox(0,0)[lt]{\lineheight{1.25}\smash{\begin{tabular}[t]{l}Snapshots \& watchpoint queries\end{tabular}}}}%
    \put(0,0){\includegraphics[width=\unitlength,page=13]{gb-state-learner.pdf}}%
  \end{picture}%
\endgroup%

%% file: io-mapping.pdf_tex
\begingroup%
  \makeatletter%
  \providecommand\color[2][]{%
    \errmessage{(Inkscape) Color is used for the text in Inkscape, but the package 'color.sty' is not loaded}%
    \renewcommand\color[2][]{}%
  }%
  \providecommand\transparent[1]{%
    \errmessage{(Inkscape) Transparency is used (non-zero) for the text in Inkscape, but the package 'transparent.sty' is not loaded}%
    \renewcommand\transparent[1]{}%
  }%
  \providecommand\rotatebox[2]{#2}%
  \newcommand*\fsize{\dimexpr\f@size pt\relax}%
  \newcommand*\lineheight[1]{\fontsize{\fsize}{#1\fsize}\selectfont}%
  \ifx\svgwidth\undefined%
    \setlength{\unitlength}{537bp}%
    \ifx\svgscale\undefined%
      \relax%
    \else%
      \setlength{\unitlength}{\unitlength * \real{\svgscale}}%
    \fi%
  \else%
    \setlength{\unitlength}{\svgwidth}%
  \fi%
  \global\let\svgwidth\undefined%
  \global\let\svgscale\undefined%
  \makeatother%
  \begin{picture}(1,0.23649906)%
    \lineheight{1}%
    \setlength\tabcolsep{0pt}%
    \put(0,0){\includegraphics[width=\unitlength,page=1]{io-mapping.pdf}}%
    \put(0.0190215,0.19953508){\color[rgb]{0,0,0}\makebox(0,0)[lt]{\lineheight{1.25}\smash{\begin{tabular}[t]{l}\pA\end{tabular}}}}%
    \put(0,0){\includegraphics[width=\unitlength,page=2]{io-mapping.pdf}}%
    \put(0.27498883,0.19953508){\color[rgb]{0,0,0}\makebox(0,0)[lt]{\lineheight{1.25}\smash{\begin{tabular}[t]{l}\pB\end{tabular}}}}%
    \put(0,0){\includegraphics[width=\unitlength,page=3]{io-mapping.pdf}}%
    \put(0.5309561,0.19953508){\color[rgb]{0,0,0}\makebox(0,0)[lt]{\lineheight{1.25}\smash{\begin{tabular}[t]{l}\pC\end{tabular}}}}%
    \put(0,0){\includegraphics[width=\unitlength,page=4]{io-mapping.pdf}}%
    \put(0.78692341,0.19953508){\color[rgb]{0,0,0}\makebox(0,0)[lt]{\lineheight{1.25}\smash{\begin{tabular}[t]{l}\pD\end{tabular}}}}%
    \put(0,0){\includegraphics[width=\unitlength,page=5]{io-mapping.pdf}}%
  \end{picture}%
\endgroup%

%% file: log-alignment.pdf_tex
\begingroup%
  \makeatletter%
  \providecommand\color[2][]{%
    \errmessage{(Inkscape) Color is used for the text in Inkscape, but the package 'color.sty' is not loaded}%
    \renewcommand\color[2][]{}%
  }%
  \providecommand\transparent[1]{%
    \errmessage{(Inkscape) Transparency is used (non-zero) for the text in Inkscape, but the package 'transparent.sty' is not loaded}%
    \renewcommand\transparent[1]{}%
  }%
  \providecommand\rotatebox[2]{#2}%
  \newcommand*\fsize{\dimexpr\f@size pt\relax}%
  \newcommand*\lineheight[1]{\fontsize{\fsize}{#1\fsize}\selectfont}%
  \ifx\svgwidth\undefined%
    \setlength{\unitlength}{465.99998474bp}%
    \ifx\svgscale\undefined%
      \relax%
    \else%
      \setlength{\unitlength}{\unitlength * \real{\svgscale}}%
    \fi%
  \else%
    \setlength{\unitlength}{\svgwidth}%
  \fi%
  \global\let\svgwidth\undefined%
  \global\let\svgscale\undefined%
  \makeatother%
  \begin{picture}(1,0.2811159)%
    \lineheight{1}%
    \setlength\tabcolsep{0pt}%
    \put(0,0){\includegraphics[width=\unitlength,page=1]{log-alignment.pdf}}%
    \put(0.01781778,0.22497274){\color[rgb]{0,0,0}\makebox(0,0)[lt]{\lineheight{1.25}\smash{\begin{tabular}[t]{l}\pA\end{tabular}}}}%
    \put(0,0){\includegraphics[width=\unitlength,page=2]{log-alignment.pdf}}%
    \put(0.34876044,0.22497274){\color[rgb]{0,0,0}\makebox(0,0)[lt]{\lineheight{1.25}\smash{\begin{tabular}[t]{l}\pB\end{tabular}}}}%
    \put(0,0){\includegraphics[width=\unitlength,page=3]{log-alignment.pdf}}%
    \put(0.67905661,0.22497274){\color[rgb]{0,0,0}\makebox(0,0)[lt]{\lineheight{1.25}\smash{\begin{tabular}[t]{l}\pC\end{tabular}}}}%
    \put(0,0){\includegraphics[width=\unitlength,page=4]{log-alignment.pdf}}%
  \end{picture}%
\endgroup%

%% file: exampleProtocolv1.pdf_tex
\begingroup%
  \makeatletter%
  \providecommand\color[2][]{%
    \errmessage{(Inkscape) Color is used for the text in Inkscape, but the package 'color.sty' is not loaded}%
    \renewcommand\color[2][]{}%
  }%
  \providecommand\transparent[1]{%
    \errmessage{(Inkscape) Transparency is used (non-zero) for the text in Inkscape, but the package 'transparent.sty' is not loaded}%
    \renewcommand\transparent[1]{}%
  }%
  \providecommand\rotatebox[2]{#2}%
  \newcommand*\fsize{\dimexpr\f@size pt\relax}%
  \newcommand*\lineheight[1]{\fontsize{\fsize}{#1\fsize}\selectfont}%
  \ifx\svgwidth\undefined%
    \setlength{\unitlength}{150bp}%
    \ifx\svgscale\undefined%
      \relax%
    \else%
      \setlength{\unitlength}{\unitlength * \real{\svgscale}}%
    \fi%
  \else%
    \setlength{\unitlength}{\svgwidth}%
  \fi%
  \global\let\svgwidth\undefined%
  \global\let\svgscale\undefined%
  \makeatother%
  \begin{picture}(1,1.04)%
    \lineheight{1}%
    \setlength\tabcolsep{0pt}%
    \put(0,0){\includegraphics[width=\unitlength,page=1]{exampleProtocolv1.pdf}}%
    \put(0.28395338,0.35101548){\color[rgb]{0,0,0}\makebox(0,0)[lt]{\lineheight{1.25}\smash{\begin{tabular}[t]{l}2\end{tabular}}}}%
    \put(0,0){\includegraphics[width=\unitlength,page=2]{exampleProtocolv1.pdf}}%
    \put(0.68137837,0.36923509){\color[rgb]{0,0,0}\makebox(0,0)[lt]{\lineheight{1.25}\smash{\begin{tabular}[t]{l}3\end{tabular}}}}%
    \put(0,0){\includegraphics[width=\unitlength,page=3]{exampleProtocolv1.pdf}}%
    \put(0.50436201,0.60736411){\color[rgb]{0,0,0}\makebox(0,0)[lt]{\lineheight{1.25}\smash{\begin{tabular}[t]{l}1\end{tabular}}}}%
    \put(0,0){\includegraphics[width=\unitlength,page=4]{exampleProtocolv1.pdf}}%
    \put(0.08813875,0.07409009){\color[rgb]{0,0,0}\makebox(0,0)[lt]{\lineheight{1.25}\smash{\begin{tabular}[t]{l}4\end{tabular}}}}%
    \put(0,0){\includegraphics[width=\unitlength,page=5]{exampleProtocolv1.pdf}}%
    \put(0.61653095,0.07408219){\color[rgb]{0,0,0}\makebox(0,0)[lt]{\lineheight{1.25}\smash{\begin{tabular}[t]{l}6\end{tabular}}}}%
    \put(0,0){\includegraphics[width=\unitlength,page=6]{exampleProtocolv1.pdf}}%
    \put(0.89689053,0.07408219){\color[rgb]{0,0,0}\makebox(0,0)[lt]{\lineheight{1.25}\smash{\begin{tabular}[t]{l}7\end{tabular}}}}%
    \put(0,0){\includegraphics[width=\unitlength,page=7]{exampleProtocolv1.pdf}}%
    \put(0.33962849,0.07409009){\color[rgb]{0,0,0}\makebox(0,0)[lt]{\lineheight{1.25}\smash{\begin{tabular}[t]{l}5\end{tabular}}}}%
    \put(0,0){\includegraphics[width=\unitlength,page=8]{exampleProtocolv1.pdf}}%
    \put(0.66380982,0.56743673){\color[rgb]{0,0,0}\makebox(0,0)[lt]{\lineheight{1.25}\smash{\begin{tabular}[t]{l}\textit{c}\textit{k}\textit{e}\textit{ }\textit{/}\end{tabular}}}}%
    \put(0.65501826,0.52743681){\color[rgb]{0,0,0}\makebox(0,0)[lt]{\lineheight{1.25}\smash{\begin{tabular}[t]{l}\textit{e}\textit{m}\textit{p}\textit{t}\textit{y}\end{tabular}}}}%
    \put(0,0){\includegraphics[width=\unitlength,page=9]{exampleProtocolv1.pdf}}%
    \put(0.26512783,0.55093241){\color[rgb]{0,0,0}\makebox(0,0)[lt]{\lineheight{1.25}\smash{\begin{tabular}[t]{l}\textit{a}\textit{p}\textit{p}\textit{d}\textit{a}\textit{t}\textit{a}\textit{ }\textit{/}\textit{ }\end{tabular}}}}%
    \put(0.28872367,0.5109325){\color[rgb]{0,0,0}\makebox(0,0)[lt]{\lineheight{1.25}\smash{\begin{tabular}[t]{l}\textit{e}\textit{m}\textit{p}\textit{t}\textit{y}\end{tabular}}}}%
    \put(0,0){\includegraphics[width=\unitlength,page=10]{exampleProtocolv1.pdf}}%
    \put(0.3894699,0.26314325){\color[rgb]{0,0,0}\makebox(0,0)[lt]{\lineheight{1.25}\smash{\begin{tabular}[t]{l}\textit{c}\textit{k}\textit{e}\textit{ }\textit{/}\textit{ }\end{tabular}}}}%
    \put(0.38067834,0.22314333){\color[rgb]{0,0,0}\makebox(0,0)[lt]{\lineheight{1.25}\smash{\begin{tabular}[t]{l}\textit{e}\textit{m}\textit{p}\textit{t}\textit{y}\end{tabular}}}}%
    \put(0,0){\includegraphics[width=\unitlength,page=11]{exampleProtocolv1.pdf}}%
    \put(0.04905972,0.28184641){\color[rgb]{0,0,0}\makebox(0,0)[lt]{\lineheight{1.25}\smash{\begin{tabular}[t]{l}\textit{a}\textit{p}\textit{p}\textit{d}\textit{a}\textit{t}\textit{a}\textit{ }\textit{/}\end{tabular}}}}%
    \put(0.07682873,0.24184649){\color[rgb]{0,0,0}\makebox(0,0)[lt]{\lineheight{1.25}\smash{\begin{tabular}[t]{l}\textit{e}\textit{m}\textit{p}\textit{t}\textit{y}\end{tabular}}}}%
    \put(0,0){\includegraphics[width=\unitlength,page=12]{exampleProtocolv1.pdf}}%
    \put(0.81048982,0.29025354){\color[rgb]{0,0,0}\makebox(0,0)[lt]{\lineheight{1.25}\smash{\begin{tabular}[t]{l}\textit{a}\textit{p}\textit{p}\textit{d}\textit{a}\textit{t}\textit{a}\textit{ }\textit{/}\end{tabular}}}}%
    \put(0.83408566,0.25025362){\color[rgb]{0,0,0}\makebox(0,0)[lt]{\lineheight{1.25}\smash{\begin{tabular}[t]{l}\textit{e}\textit{m}\textit{p}\textit{t}\textit{y}\end{tabular}}}}%
    \put(0,0){\includegraphics[width=\unitlength,page=13]{exampleProtocolv1.pdf}}%
    \put(0.52072566,0.27224348){\color[rgb]{0,0,0}\makebox(0,0)[lt]{\lineheight{1.25}\smash{\begin{tabular}[t]{l}\textit{c}\textit{k}\textit{e}\textit{ }\textit{/}\end{tabular}}}}%
    \put(0.50915525,0.23224356){\color[rgb]{0,0,0}\makebox(0,0)[lt]{\lineheight{1.25}\smash{\begin{tabular}[t]{l}\textit{c}\textit{l}\textit{o}\textit{s}\textit{e}\textit{d}\end{tabular}}}}%
  \end{picture}%
\endgroup%

%% file: st-update-single-column.pdf_tex
\begingroup%
  \makeatletter%
  \providecommand\color[2][]{%
    \errmessage{(Inkscape) Color is used for the text in Inkscape, but the package 'color.sty' is not loaded}%
    \renewcommand\color[2][]{}%
  }%
  \providecommand\transparent[1]{%
    \errmessage{(Inkscape) Transparency is used (non-zero) for the text in Inkscape, but the package 'transparent.sty' is not loaded}%
    \renewcommand\transparent[1]{}%
  }%
  \providecommand\rotatebox[2]{#2}%
  \newcommand*\fsize{\dimexpr\f@size pt\relax}%
  \newcommand*\lineheight[1]{\fontsize{\fsize}{#1\fsize}\selectfont}%
  \ifx\svgwidth\undefined%
    \setlength{\unitlength}{408.99998474bp}%
    \ifx\svgscale\undefined%
      \relax%
    \else%
      \setlength{\unitlength}{\unitlength * \real{\svgscale}}%
    \fi%
  \else%
    \setlength{\unitlength}{\svgwidth}%
  \fi%
  \global\let\svgwidth\undefined%
  \global\let\svgscale\undefined%
  \makeatother%
  \begin{picture}(1,0.59902201)%
    \lineheight{1}%
    \setlength\tabcolsep{0pt}%
    \put(0,0){\includegraphics[width=\unitlength,page=1]{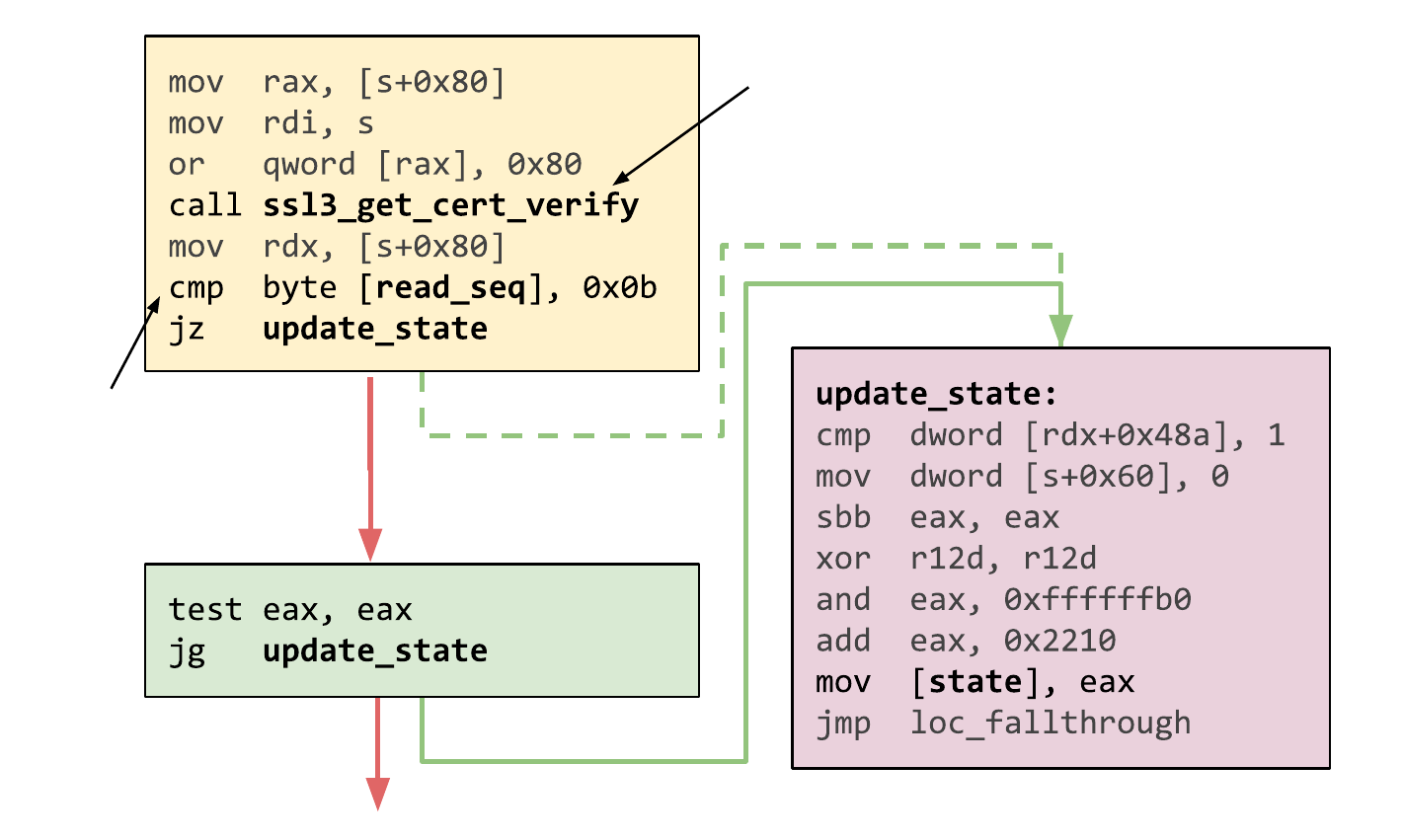}}%
    \put(0.05882406,0.28177678){\color[rgb]{0,0,0}\makebox(0,0)[lt]{\lineheight{1.25}\smash{\begin{tabular}[t]{l}\pB check \end{tabular}}}}%
    \put(0.03235505,0.25243699){\color[rgb]{0,0,0}\makebox(0,0)[lt]{\lineheight{1.25}\smash{\begin{tabular}[t]{l}backdoor depth\end{tabular}}}}%
    \put(0,0){\includegraphics[width=\unitlength,page=2]{drawing.pdf}}%
    \put(0.29845872,0.25292791){\color[rgb]{0,0,0}\makebox(0,0)[lt]{\lineheight{1.25}\smash{\begin{tabular}[t]{l}\pC skip check of \end{tabular}}}}%
    \put(0.32493967,0.22358812){\color[rgb]{0,0,0}\makebox(0,0)[lt]{\lineheight{1.25}\smash{\begin{tabular}[t]{l}verify result\end{tabular}}}}%
    \put(0,0){\includegraphics[width=\unitlength,page=3]{drawing.pdf}}%
    \put(0.78825093,0.40550351){\color[rgb]{0,0,0}\makebox(0,0)[lt]{\lineheight{1.25}\smash{\begin{tabular}[t]{l}\pD perform state \end{tabular}}}}%
    \put(0.83984049,0.37616372){\color[rgb]{0,0,0}\makebox(0,0)[lt]{\lineheight{1.25}\smash{\begin{tabular}[t]{l}change\end{tabular}}}}%
    \put(0,0){\includegraphics[width=\unitlength,page=4]{drawing.pdf}}%
    \put(0.56058393,0.54424667){\color[rgb]{0,0,0}\makebox(0,0)[lt]{\lineheight{1.25}\smash{\begin{tabular}[t]{l}\pA verify result \end{tabular}}}}%
    \put(0.57823964,0.51490687){\color[rgb]{0,0,0}\makebox(0,0)[lt]{\lineheight{1.25}\smash{\begin{tabular}[t]{l}stored in eax\end{tabular}}}}%
    \put(0,0){\includegraphics[width=\unitlength,page=5]{drawing.pdf}}%
    \put(0.08491292,0.06798032){\color[rgb]{0,0,0}\makebox(0,0)[lt]{\lineheight{1.25}\smash{\begin{tabular}[t]{l}Close\end{tabular}}}}%
    \put(0.0564092,0.03864052){\color[rgb]{0,0,0}\makebox(0,0)[lt]{\lineheight{1.25}\smash{\begin{tabular}[t]{l}Connection\end{tabular}}}}%
  \end{picture}%
\endgroup%